\documentclass[
    ,draft            
    ,sort&compress    
 ]
  {aipproc}

\layoutstyle{6x9}

\def\text#1{\rm #1}
\def\ve#1{{\bf #1}} 
\def\hve#1{{\bf #1}} 

\newcommand\dtot{{\mathrm d}}	

\newcommand\vv{{\mathrm v}}	

\newcommand\sss[1]{\bigbreak{\noindent\bf #1}\par%
                \nobreak\medskip\noindent\ignorespaces}

\begin{document}

\title{Chromospheric Dynamics and Line Formation}


\author{R.~Hammer}{
  address={Kiepenheuer-Institut f\"ur Sonnenphysik, Sch\"oneckstr.~6,
  79104 Freiburg, Germany},
  ,email={hammer@kis.uni-freiburg.de}
}

\author{P.~Ulmschneider}{
  address={Institut f\"ur Theoretische Astrophysik, Universit\"at
  Heidelberg, Albert \"Uberlestr.~2, \\
  69120 Heidelberg, Germany},
  ,email={ulmschneider@ita.uni-heidelberg.de}
}

\begin{abstract}
The solar chromosphere is very dynamic, due to the presence of large
amplitude hydrodynamic waves.  Their propagation is affected by NLTE
radiative transport in strong spectral lines, which can in turn be
used to diagnose the dynamics of the chromosphere.  We give a basic
introduction into the equations of NLTE radiation hydrodynamics and
describe how they are solved in current numerical simulations.  The
comparison with observation shows that one-dimensional codes can
describe strong brightenings quite well, but the overall chromospheric
dynamics appears to be governed by three-dimensional shock propagation.
\end{abstract}

\maketitle



\section{Introduction}

At low spatial and temporal resolution, the solar atmosphere has a
well-defined \emph{average} temperature structure
(\citet{VAL81,FAL93}; for snapshots of the most recent variants of
such models see also \citep{A07,FBH07}): from the surface (i.e., the
layer seen in continuum radiation in the visible part of the spectrum)
the temperature drops continuously within the \emph{photosphere}, from
around 5\,800\,K down to a minimum value of the order of 4\,000\,K at
a height of about 500\,km.  Beyond the minimum, the average
temperature increases first gently in the \emph{chromosphere}, then
rapidly in the \emph{transition region}, which starts at a height of
some 2\,000\,km, reaching nearly a million K even in the coolest parts
of the \emph{corona}, and several million K in the hottest parts. (For
an extended review of the solar atmosphere see e.g.\ \citet{SH02}.)  This
outward temperature rise must be supported by mechanical heating due
to a combination of magnetic and nonmagnetic mechanisms (as reviewed
e.g.\ by \citet{NU90,NU96}; \citet{UM03}), which are ultimately powered by the
convective motions below the photosphere.

\hbox to \hsize{%
{\vbox to 0pt{\vskip-18cm
    \nointerlineskip\frenchspacing\footnotesize%
    \hfill\hbox to 9.6cm{\vbox{
       \hbox{to appear in the proceedings of the 
             \emph{Kodai School on Solar Physics,}}%
       \hbox{held at Kodaikanal Observatory, Dec 10--22, 2006}%
       \hbox{eds.\ S.S.\ Hasan and D.\ Banerjee, AIP Conf.\ Procs.\ \textbf{919},}%
       \hbox{http://proceedings.aip.org/proceedings/confproceed/919.jsp}%
    \vss}}}}}\vglue-\baselineskip%
\month=5 \day=25

With the advent of more advanced instrumentation, the achievable
spatial and temporal resolution has been enhanced, and we have
increasingly realized that this crude picture can serve only as a
rough guide to the average solar atmosphere.  The real atmosphere is
characterized by a high level of fine structure, which in the upper
chromosphere (and even more in the corona) is mostly due to thermal
effects within magnetically confined plasma (\citep{K53, J06}).
Moreover, the chromosphere is extremely dynamic.  For example, much of
the emission from the upper chromosphere comes from fine structures
like spicules and fibrils (e.g., \citep{HN03, HN05, R07, DePont07}),
in which the plasma is accelerated to high velocities, up to several
times sound speed.

Even outside of active regions, the solar surface is permeated by
magnetic fields.  Stronger, long living magnetic flux concentrations
tend to be arranged by the supergranular flow in a network-like
pattern, which is visible up into the transition region.  The interior
parts of cells in this \emph{chromospheric network} are not entirely
field-free, but the magnetic field is weaker than in the network and
very likely unimportant for the dynamics.  If, and to what extent, it
affects the heating of the plasma is still under discussion.  The
importance of the magnetic field changes with height, since magnetic
flux concentrations expand with height and ultimately fill all
available space in the upper chromosphere and corona.  The current
paper will concentrate on the dynamics of these cell interior parts of
the solar chromosphere; therefore we neglect the effects of the
magnetic field in the equations to be derived in subsequent sections.

The connection between the solar chromosphere and the underlying
photosphere has recently been studied by \citet{W06}, who also
produced a movie clip \citep{W06b} that shows the photospheric and
chromospheric dynamics both in the network and in the cell interior at
high spatial resolution.

Even those parts of the solar chromosphere where magnetic structuring
is unimportant are known to be permeated by strong waves.  They
develop out of waves of relatively small amplitude that arise quite
naturally in the underlying convective layer.  When these waves travel
upward, they experience a decrease in density $\rho$, due to the
gravitational stratification.  This leads to an increase of the wave
amplitude $\vv_\mathrm{max}$, since the wave energy flux
$F_\mathrm{wave}\propto\rho \vv_\mathrm{max}^2 c_s$, where $c_s$ is the
sound speed, is conserved as long as the waves do not dissipate and if
they are not damped by radiative energy exchange between wave peaks
and valleys.  The larger the wave amplitude becomes, the more
important are nonlinear effects.  As a result of the latter, the wave
peaks propagate faster than the valleys and try to overtake them; thus
the waves ``break'' and form quasi-discontinuous shocks, which
dissipate the wave energy by thermal conduction and viscosity in their
steep shock fronts (e.g., \cite{LL59}).

Most of the solar radiation in the visible originates from the
photosphere; the chromosphere is virtually transparent at these
wavelengths.  For ground-based observations of the chromosphere one is
restricted to the central regions of a few very strong absorption
lines in the visible, near UV, and near IR; while from space one can
observe the chromosphere also in EUV emission lines and continua.  The
transition region and corona, finally, emit predominantly in the EUV,
X-ray and radio ranges.

The most important spectral lines for studying chromospheric dynamics
are the H- and K- lines as well as the infrared triplet lines of
singly ionized calcium (Ca\,{\sc ii}).  The h- and k-lines of Mg\,{\sc
ii} would also be ideal for this purpose, but they lie too far in the
UV to be observable from the ground.  These strong lines are not only
of high diagnostic value, but (along with numerous iron [mostly
Fe\,{\sc ii}] and some hydrogen lines) they represent also the
dominant cooling agents of the solar chromosphere (\citet{AA89})
and must therefore be treated adequately in numerical simulations.

\begin{figure}
\includegraphics[angle=270,width=\textwidth]{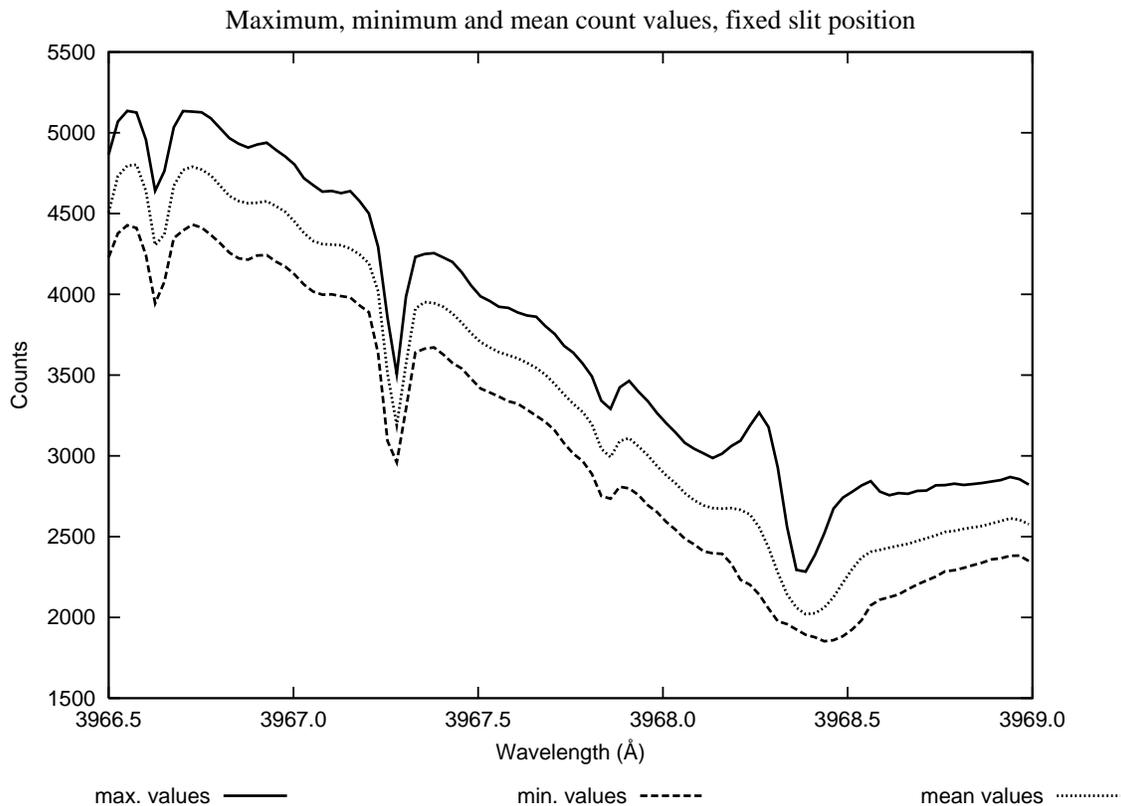} 
\caption{Inner blue wing of the very strong Ca\,{\sc ii}\,H line.  The
middle curve (dotted) shows the time-averaged profile during a
4\,300\,s observation sequence.  Each part of the line profile probes
different parts of the solar atmosphere, from the mid-chromosphere in
the line core at 3968.49\,\AA\ down to the lower photosphere in the
far line wing (not shown).  The indicated spectral range contains also
a number of weak Fe\,{\sc i} lines, which are formed in the
mid-photosphere.  When waves propagate through the atmosphere, they
affect the intensity at the respective wavelengths.  The upper and lower
curve show the maximum and minimum values that were reached during the
observation sequence.}
\label{huFig1}
\end{figure}

Fig.~\ref{huFig1} was derived from a time series of the Ca\,{\sc ii}\,H
line profile obtained in August 2005 with the Echelle spectrograph of
the German VTT telescope at the Observatorio de Tenerife
\citep{RSH07}.  As a result of the dynamic behavior of the
chromosphere during the time series, the line profile is highly
variable: Perturbations are seen all the time - sometimes small ones
moving towards the core, but more commonly one notices large parts of
a line wing change synchronously.  A movie of the line profile during
the observation demonstrates these changes (\citet{RSHK07}).
Fig.~\ref{huFig1} shows the time-averaged profile and the minimum and
maximum intensity values that occurred during the sequence.

In these lectures, we introduce the physical principles that govern
the highly dynamic behavior of solar and stellar chromospheres,
thereby restricting ourselves to those regions where magnetic fields
do not dominate the dynamics.  These physical principles are described
by the time-dependent equations of radiation hydrodynamics.  To
outline their theoretical foundation, we first derive the basic
hydrodynamic and thermodynamic equations.  Next we discuss elements of
radiation theory, first the basic concepts and then the complications
that arise in chromospheres.  Finally we give a brief overview of how
these equations are solved in numerical calculations and to what
extent current state-of-the-art simulations can describe the observed
chromospheric dynamics.  All main chapters are preceded by
recommendations of literature suitable for further reading.


\section{Basic hydrodynamics and thermodynamics}
General references on the hydrodynamical, thermodynamical, and
mathematical concepts treated in this chapter include \citep{LL59,
HCB64, S62, S69, MF53}.

\subsection{Continuity Equation, Euler Frame} 
 
\begin{figure}[h]
\includegraphics[height=3.5cm]{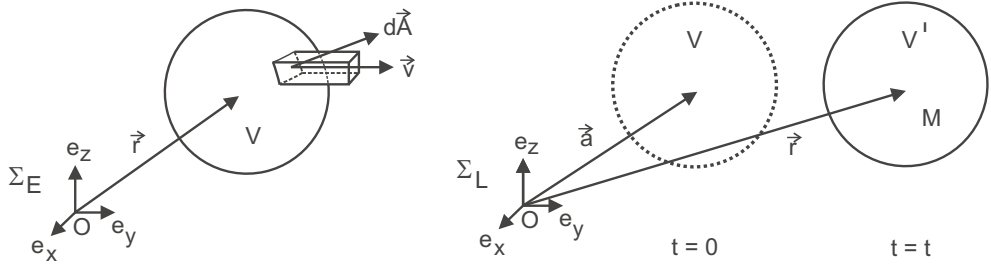}
\caption{Euler frame $\Sigma_E$ (left): observe the mass flow out of a 
fixed volume $V$; 
Lagrange frame $\Sigma_L$ (right): follow a mass element $M$}
\label{huFig2}
\end{figure}

In the {\it Euler frame} $\Sigma_E$ 
we consider a Cartesian coordinate system 
with origin $O$ and unit vectors $\hve e_x$, $\hve e_y$, $\hve
e_z$ in $x$-, $y$-, $z$-directions.  We observe the density
$\rho(\ve{r},t)$ [\,g\,cm$^{-3}$\,]\footnote{We consider it
instructive to specify, in
square brackets, units for newly introduced quantities.  This proves
to be particularly helpful for radiative terms introduced below.  
We choose CGS units.}, 
gas pressure $p(\ve{r},t)$ [\,dyn\,cm$^{-2}$\,],
and temperature $T(\ve{r},t)$ [\,K\,] of plasma flowing out of 
a fixed volume $V$ with velocity $\ve{v}(\ve{r},t)$ [\,cm\,s$^{-1}$\,] 
(Fig.~\ref{huFig2}).
$\ve{r}$ is the radius vector with components $(x,y,z)$, and $t$ is the time.
Let d$\ve A$ be a directed surface element of $V$. 
Then the amount of matter
flowing out of $V$ across its surface per unit time (Fig.~\ref{huFig2}) is
\begin{equation}
\oint \rho \ve v \cdot \dtot \ve A =\int\limits_V \nabla 
\cdot \rho  
\ve{v}\ \dtot V \ \ \ , 
\label{e1.1}
\end{equation}
where the equal sign is due to Gauss's theorem. The amount
of matter missing per unit time
from volume $V$ is given by
\begin{equation}
-{\partial \over{\partial 
t}} \int\limits_V \rho  \dtot V= -\int\limits_V {\partial \rho 
\over{\partial t}}\ \dtot V 
\ \ \ ,
\label{e1.2}
\end{equation} 
where the equal sign is due to the fact 
that $V$ does 
not depend on time.  Equating Eqs.~(\ref{e1.1}), (\ref{e1.2}), 
and because $V$ 
is arbitrary one finds the 
{\it continuity equation}: 
\begin{equation}
 {\partial 
\rho \over {\partial  t}}+ \nabla \cdot \rho \ve 
v =0 
\ \ \ .
\label{e1.3}
\end{equation}

\subsection{{Lagrange Frame}}
In a \emph{Lagrange frame} $\Sigma_L$ we monitor the mass element $M$ 
(Fig.~\ref{huFig2}) initially contained in $V$ (at time $t=0$ and position $\ve r=
\ve a$), which at a later time $t$ is in volume $V'$ at radius vector
$\ve r(\ve a,t)=\left(x(\ve a,t),y(\ve a,t),z(\ve a,t)\right)$. Vector
$\ve a$ uniquely identifies a mass element at the initial time $t=0$.
We consider physical variables such as $\rho$ on an (x,t)-plane, 
on which the position of the mass element $M$ describes 
a path $x(t)$, where $x(t=0)=a$, so that $\rho = \rho (x(t),t)$. 
The rate of change of
$\rho$ along that path can be pictured as the sum of the
incremental variations of $\rho $ in $t$- and $x$-directions,
\begin{equation}
{\dtot \rho \over {\dtot t}} \equiv 
\left({\partial \rho \over {\partial t}}\right)_a = 
\left({\partial \rho \over {\partial t}}\right)_x + 
\left( {\partial \rho \over {\partial x}}\right)_t {\dtot x(t)\over{ \dtot t}}
= \left({\partial \rho \over {\partial t }}\right)_x + 
\vv_x \left({\partial \rho \over {\partial x  }}\right)_t\ \ \ , 
\label{e1.4}
\end{equation}
where the sign ($\equiv$) means equal by definition and
\begin{equation}
\vv_x = {\dtot x(t)\over \dtot t}\ \ \ 
\label{e1.5}
\end{equation}  
is the velocity of the mass element in $x$-direction. The time derivative 
$\dtot/\dtot t \equiv \left({\partial /{\partial t}}\right)_a$ 
in the Lagrange frame is the so called {\it substantial derivative}, 
sometimes also called \emph{material} or \emph{total} derivative. 
It describes the time-dependence of
a physical quantity in a moving mass element.  In three dimensions, and
for any physical function $h$ (such as $\rho$, $p$, $T$, etc.), the
substantial derivative is analogously given by
\begin{equation}
{\dtot h\over \dtot t}={\partial  h\over \partial t} +\ve 
v \cdot \nabla h 
\ \ \ .
\label{e1.6}
\end{equation}

\subsection{{Equation of Motion}} 
In the Lagrange frame $\Sigma_L$, according to Newton's third law the
inertial force (= mass $\dtot M$ times 
acceleration $\dtot\ve v / \dtot t$) is balanced
by the forces acting on the mass element,
\begin{equation}
\int\limits_M {\dtot \ve v\over \dtot t}\ \dtot M = 
\int\limits_V \rho {\dtot \ve v \over {\dtot t}}  \ \dtot V
= \int\limits_V \ve f\ \dtot V -\int\limits_V \nabla p\ \dtot V
\ \ \ . 
\label{e1.7}
\end{equation} 
Here the acting forces consist of a volume force density
$\ve f$ [\,dyn\,cm$^{-3}$\,] and a pressure force. In 
stellar atmospheres the volume force density is due to gravity, 
\begin{equation}
\ve f = - {G\ M_\star \ \rho \over r^2}\ 
\hve{e}_r 
= - \rho\ g\ \hve{e}_r
\ \ \ , 
\label{e1.8}
\end{equation} 
 where $\hve{e}_r$ is the unit vector in radial direction,
$M_\star $ the stellar mass, 
$G = 6.674\times 10^{- 8}$ cm$^3$\,g$^{-1}$\,s$^{-2}$ 
the gravitational constant, and $r$ the radial distance from
the center of the star.  As the extent $\Delta r$ of stellar atmospheres
is often small ($\Delta r \ll r$), we can replace $G M_\star /r^2$ by a
constant gravitational acceleration $g$, e.g.\ $g= 2.74\times 10^4$
cm\,s$^{-2}$ for the solar atmosphere. Using Eqs.~(\ref{e1.6}),
(\ref{e1.7}), and (\ref{e1.8}),
and noting that
the volume of the mass element is arbitrary, we obtain the {\it equation
of motion}
\begin{equation}
\rho \left({\partial \ve v \over{\partial t}} 
+\ve v \cdot 
\nabla 
\ve v \right)=-\nabla p-\rho g\ \hve{e}_r\ \ \ . 
\label{e1.9}
\end{equation} 
In this equation we have neglected {\it viscosity} and {\it radiation
pressure}.  It can be shown that both are unimportant in the
atmospheres of typical late-type stars (i.e., stars of spectral type
beyond late A).

\subsection{{Energy Equation}}
It is convenient to picture the moving 
mass element 
as enclosed by a wall and
to carefully monitor the energy flowing
through that wall.
A powerful book-keeping quantity for this is the 
specific (i.e., per unit mass) entropy 
$S$ [\,erg\,g$^{-1}$\,K$^{-1}$\,].
If the walls are impermeable, energy conservation 
just means
$S$ = constant in $\Sigma_L$. This can be written 
with Eq.~(\ref{e1.6})
\begin{equation}
{\dtot S\over \dtot t}={\partial S\over \partial t} +\ve 
v \cdot  \nabla
S = 0\ \ \ .  
\label{e1.10}
\end{equation}
In atmospheric regions close to the 
star (in photospheres and most of the
chromospheres) only two processes
by which mass elements gain energy are important,
{\it radiation} and {\it Joule heating}.  In coronae 
and some chromospheric regions one 
has steep temperature gradients
and sometimes large velocity gradients. Here 
{\it thermal 
conduction} and {\it viscosity} are other important heating (or 
energy loss) mechanisms. 
One thus has the relation
\begin{equation} 
{\dtot S\over \dtot t}={\partial S \over \partial t}
+\ve{v} \cdot \nabla S = \left. {\dtot S \over \dtot t}
\right \vert_\mathrm{ext} = 
 {\Phi_\mathrm{R} \over \rho T} + {\Phi_\mathrm{J} \over \rho T} + 
{\Phi_\mathrm{C} \over \rho T} 
+ {\Phi_\mathrm{V} \over \rho T} \ \ \ , 
\label{e1.11}
\end{equation} 
called {\it entropy conservation equation}.  Here $\left. \dtot S/
\dtot t \right \vert_\mathrm{ext}$ [\,erg\,g$^{-1}$\,K$^{-1}$\,s$^{-1}$\,] is
the heating function resulting from external heating by radiative,
Joule, thermal conductive, and viscous heating, where
$\Phi_\mathrm{R}$, $\Phi_\mathrm{C}$, and $\Phi_\mathrm{V}$
[\,erg\,cm$^{-3}$\,s$^{-1}$\,] are the net radiative, thermal
conductive, and viscous heating rates, respectively.  These functions
will be discussed below. Joule heating, $\Phi_\mathrm{J}$, will not be
considered here, although magnetic fields and associated heating
(magnetohydrodynamics, MHD) are important at least in some parts of
stellar chromospheres, as discussed above.

\subsection{Atmospheric Gas Composition}
Stellar atmospheres consist of ideal 
gases, which are 
described by the universal gas law
\begin{equation} 
p= \sum_i n_i k T = \rho {\Re T \over \mu}
\ \ \ ,  
\label{e1.12}
\end{equation}
where $k=1.3807\times 10^{-16}$ erg\,K$^{-1}$ is the Boltzmann
constant, $\mu$ [\,g\,mol$^{-1}$\,] the mean molecular weight, and
$\Re= 8.3145\times 10^7$ erg\,K$^{-1}$\,mol$^{-1}$ the
universal gas constant. 
$n_i$ are the number densities [\,cm$^{-3}$\,] of the different types of
particles $i$ (atoms, ions, and electrons).  With a given mixture of chemical
elements in the stellar gas the mean molecular weight 
can be computed as a function $\mu=\mu(T,p)$.
For solar type neutral gas $ \mu \approx 1.24$, while for fully ionized gas
$\mu \approx 0.60$.

\subsection{Basic Elements of Thermodynamics}
In many applications in stellar 
photospheres, 
chromospheres, and coronae it is 
sufficient to consider atmospheric gases that
are 
either neutral or fully ionized. Under these 
conditions the 
thermodynamic relations are particularly simple. 
The specific internal energy $E_s$ [\,erg\,g$^{-1}$\,] is 
given by
\begin{equation} 
E_s=c_v T\ \ \ ,  
\label{e1.14}
\end{equation}
where $c_v$ [\,erg\,g$^{-1}$\,K$^{-1}$\,] is 
the specific heat per unit mass for 
constant volume, given by
\begin{equation} 
c_v= {1\over \gamma-1} {\Re \over \mu}\ \ \ , 
\label{e1.15}
\end{equation} 
and $\gamma=c_p/c_v$ is the ratio 
of specific heats, which is constant ($\gamma=5/3$) for either neutral
or fully ionized gases. 
With the specific volume $V_s$ [\,cm$^3$\,g$^{-1}$\,],
\begin{equation} 
V_s={1\over \rho}\ \ \ ,
\label{e1.16}
\end{equation}
and the fundamental laws of thermodynamics we obtain
\begin{equation} 
T\ \dtot S= \dtot E_s+p\ \dtot V_s=\dtot E_s-{p\over\rho^2}  \ \dtot \rho\ \ 
\ . 
\label{e1.17}
\end{equation}
From these equations the relationships between the thermodynamical 
variables $S$, $E_s$, $T$, $p$, and $\rho$
can be computed.  Specification of two of these allows to determine
the remaining variables.

\subsection{{Conservation Equations}}
It is often convenient to write the 
hydrodynamic
equations in conservation form, in terms of
conserved quantities $f$
(mass, momentum, energy) and associated fluxes $\ve F$ (mass 
flux,
momentum flux, energy flux):
\begin{equation} 
{\partial f\over \partial t} =-\nabla\cdot \ve 
F + C\ \ \ ,
\label{e1.24}
\end{equation} 
where $C$ is a source term.

\sss{a. Conservation of mass} 
The continuity equation (\ref{e1.3}) is already in 
conservation form:
\begin{equation}
{\partial \rho \over \partial t} =-\nabla \cdot 
\rho \ve v
\ \ \ . 
\label{e1.25}
\end{equation}

\sss{b. Conservation of momentum}
From Eqs.~(\ref{e1.3}) and (\ref{e1.9}) one obtains
\begin{equation}
{\partial \rho \ve v \over \partial t} =\rho 
{\partial \ve v
\over  \partial t}+\ve v {\partial \rho \over 
\partial t}
= -\rho \ve v\cdot \nabla \ve v -\ve v \nabla 
\cdot \rho
\ve v -\nabla p- \rho g \hve{e}_r\ \ \ , 
\label{e1.26}
\end{equation} 
which can be written
\begin{equation}
{\partial \rho \ve v \over{\partial t}} =-
\nabla \cdot
\left(\rho \ve v \ve v +p {\overline{\overline 
U}}\right) -\rho
 g \hve{e}_r\ \ \ . 
\label{e1.27}
\end{equation} 
Here $\ve v \ve v$ is a dyad (e.g., \citep{S69, MF53}), and   
\begin{equation}
{\overline {\overline U}}= \hve{e}_x 
\hve{e}_x+\hve{e} _y \hve{e} _y+
\hve{e}_z \hve{e}_z\ \ \ ,
\label{e1.28}
\end{equation}  
is a unit dyad.

\sss{c. Conservation of energy}
There are three types of energy densities [\,erg\,cm$^{- 3}$\,]
for gas elements in nonmagnetic stellar atmospheres:

$\rho E_s\ \ $ internal energy (energy of microscopic undirected motion of
atoms and ions),

${1\over 2} \rho \vv^2\  $ kinetic energy (directed motion of the
entire gas element),

$\rho \phi\ \ $ potential energy (gas element in the gravitational
field).

\noindent The gravitational potential is given by 
\begin{equation} 
\phi \equiv -
{G M_\star  \over r}\ \ \ . 
\label{e1.29}
\end{equation}
With this and Eq.~(\ref{e1.8}) the volume force density 
can be written
\begin{equation}
\ve f=-\rho g \hve{e}_r =-\rho {G M_\star \over 
{r^2}} \hve{e}_r=-
\rho \nabla \phi\ \ \ .   
\label{e1.30}
\end{equation}
The time derivative of the total energy density
can be written as a sum of four terms
\begin{equation} 
{\partial \over \partial t} \left( {1 \over 2} 
\rho\vv^2+\rho E_s+\rho \phi \right) = 
{1\over 2} \vv^2 {\partial
\rho \over \partial t} + 
\rho \ve v\cdot {\partial \ve v \over \partial t} + 
{\partial \rho E_s\over \partial t} + 
{\partial \rho \phi \over \partial t}\ \ \ . 
\label{e1.31}
\end{equation} 
Modifying the terms on the RHS using Eqs.~(\ref{e1.3}), (\ref{e1.9}),
(\ref{e1.11}), (\ref{e1.17}), and (\ref{e1.30}), one obtains the {\it 
energy conservation equation}: 
\[
 {\partial \over \partial t} 
\left({1\over 2} \rho \vv^2 +\rho E_s +\rho \phi 
\right) =-\nabla 
\cdot \rho \ve v \left({1\over 2} \vv^2+E_s+{p\over 
\rho}+\phi 
\right) +\rho T \left. {\dtot S\over \dtot t} \right| 
_\mathrm{ext} \ \ \ 
\]
\begin{equation}
=-\nabla 
\cdot \rho \ve v \left({1\over 2} \vv^2+E_s+{p\over 
\rho}+\phi 
\right) + 
 \Phi_\mathrm{R}  + \Phi_\mathrm{J}  + \Phi_\mathrm{C} + \Phi_\mathrm{V}  \ \ \ . 
\label{e1.39}
\end{equation} 
It is seen that there are three energy flux components 
[\,erg\,cm$^{-2}$\,s$^{-1}$\,]:

$\rho \ve v {1\over 2} \vv^2\ \ $  kinetic energy 
flux,

$\rho \ve v\left(E_s+ p/\rho \right)\ \ $ enthalpy 
flux,

$\rho \ve v \phi \ \ $ potential energy flux.

\subsection{Heating by Viscosity and Thermal Conductivity}
The particle transport processes which 
occur in the presence of 
velocity and temperature gradients contribute to 
the local 
heating.

\noindent
The thermal conductive heating rate $\Phi_\mathrm{C}$  
[\,erg\,cm$^{-3}$\,s$^{-1}$\,] is 
given by
\begin{equation}
\Phi_\mathrm{C} ={\dtot \over \dtot x} \kappa_{th} {\dtot T \over \dtot x}\ \ \ .
\label{e1.59}
\end{equation}
The viscous heating rate $\Phi_\mathrm{V}$ [\,erg\,cm$^{-3}$\,s$^{-1}$\,] is
\begin{equation}
\Phi_\mathrm{V} = \eta_{vis} \left( {\dtot v \over \dtot x} \right)^2\ \ \ .
\label{e1.60}
\end{equation} 
The coefficients of thermal conductivity $\kappa_{th}$ and
viscosity $\eta_{vis}$ are functions of $T$ and $p$.


\section{Elementary Radiation Theory}
Using the three time-dependent hydrodynamic equations (\ref{e1.25}), 
(\ref{e1.27}), and (\ref{e1.39}) together with the ideal gas law Eq.~(\ref{e1.12}), the 
thermodynamic relations and functions such as $\mu(T,p)$, $\kappa_{th}(T,p)$ and
$\eta_{vis}(T,p)$, one would be able to compute the dynamics of stellar
chromospheres, except that radiation is critically important.  For
this reason we now give a short review of the basic equations of
radiation theory and derive the radiative heating rate.  For further
reading we suggest in particular \citep{M78, R03}, but also
\citep{S02, S03}, and for looking up equations and 
atomic data \citep{A73, C00}.

\subsection{Basic Radiation Quantities}
\begin{figure}[h]
\includegraphics[height=4.5cm]{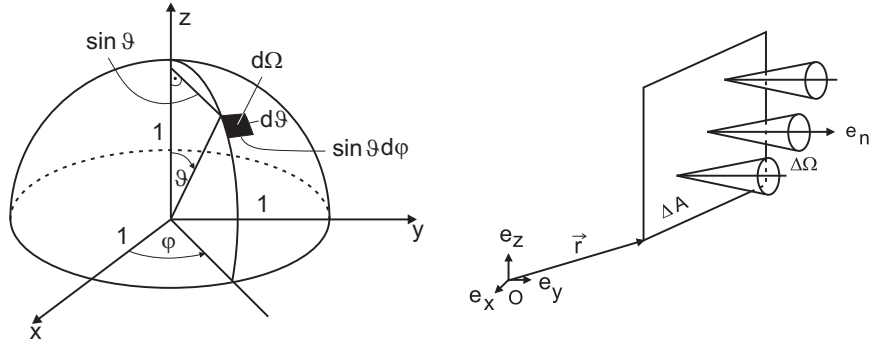}
\caption{Left: Solid angle. Right: Intensity}
\label{huFig3}
\end{figure}

\sss{a. Solid angle}
Consider an orthogonal Cartesian coordinate system 
$(x,y,z)$. Let the latitude angle $\vartheta$ and azimuth angle 
$\varphi$ be defined as seen in Fig.~\ref{huFig3}, left. The {\it solid angle} 
is a surface element on a sphere with unit radius
\begin{equation}
 \dtot \Omega \equiv \sin \vartheta\ \dtot \vartheta\ \dtot \varphi\ \ \ ,
\label{e3.1}
\end{equation} 
and has the dimension [\,sr\,] (for \emph{steradian}).

\sss{b. Intensity}
The intensity is the basic quantity of  
radiation theory.  Consider the energy
$\Delta  E$ of all photons that flow through 
a window of area $\Delta A$ in normal
direction $\ve e_n$ into the solid angle
$\Delta \Omega$ per time interval 
$\Delta t$ and frequency interval
$\Delta \nu$ (Fig.~\ref{huFig3}, right).  Assume that $\ve r$ is the radius 
vector that points from the origin $O$ of a coordinate
system (with unit vectors $\ve e_x$, $\ve e_y$, $\ve e_z$) 
to the location of the area $\Delta \ve A$.  Then
the {\it monochromatic intensity} $I_\nu$, also called
{\it specific intensity}, is given by
\begin{equation}
 I _\nu \left( \ve r,\ve e_n,t \right) \equiv
\lim {\Delta E \over{ \Delta A\ \Delta t 
\ \Delta\Omega\ \Delta \nu}}\ \ \ . 
\label{e3.2}
\end{equation}
The limit is taken for $\Delta A, \Delta t ,
\Delta \Omega , \Delta \nu \rightarrow 0$. $I_\nu$ has the dimension 
[\,erg\,cm$^{-2}$\,s$^{-1}$\,sr$^{-1}$ Hz$^{-1}$\,] 
= 10$^{-3}$ [\,W\,m$^{-2}$\,sr$^{-1}$\,Hz$^{-1}$\,]. Recall that
1 J = 1 Ws = 1 Nm = 1 kg\,m$^2$\,s$^{-2}$ = 10$^7$ erg.

\sss{c. Mean intensity }
The {\it mean intensity} is the specific intensity 
averaged over all angles,
\begin{equation}
J_\nu \left( \ve r, t\right) ={1\over {4\pi}}
\oint I_\nu\left( \ve r,\ve e_n ,t \right)\ \dtot \Omega=
{1\over{4 \pi}}{\int\limits_0^\pi}
{\int\limits_0^{2 \pi}} I_\nu\ \sin \vartheta\ \dtot \vartheta
\ \dtot\varphi\ \ \ . 
\label{e3.3}
\end{equation}  
In most cases there is no dependence of $I_\nu$ 
on $\varphi$, and with the {\it angle cosine} 
\begin{equation}
\mu \equiv \cos \vartheta\ \ \ 
\label{e3.4}
\end{equation}
the mean intensity can be written
\begin{equation}
 J_\nu\left( \ve r,t \right) = {1\over 2}{\int\limits
_0^\pi} I_\nu\ {\underbrace{\sin\vartheta\ \dtot \vartheta}
_{-\dtot \mu}} ={1\over 2} {\int \limits _{-1} ^{+1}} 
I_\nu\ \dtot \mu\ \ \ . 
\label{e3.5}
\end{equation} 
The dimension of $J_\nu$, 
[\,erg\,cm$^{-2}$\,s$^{-1}$\,sr$^{-1}$\,Hz$^{-1}$\,], 
is the same as that of $I_\nu$.

\begin{figure}[t]
\includegraphics[height=4.4cm]{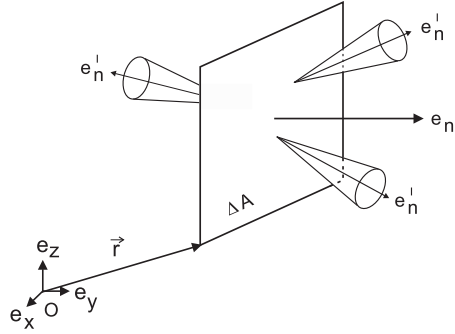}
\caption{Radiative flux}
\label{huFig4}
\end{figure}

\sss{d. Radiative flux}   
Consider a window
with area $\Delta A$ and normal $\ve e_n$ (Fig.~\ref{huFig4}). 
Assume that from both sides photons flow in arbitrary 
directions $\ve e_n'$ through  $\Delta \ve A$. The net energy
$\Delta E$ transported in  
$\ve e_n$ direction by all the photons flowing
through $\Delta A$ per time interval $\Delta t$
and frequency band $\Delta \nu$ is the {\it radiative flux} given by
\begin{equation}
 F_ \nu \left(\ve r,\ve e_n ,t\right) \ 
= \lim {\Delta E \over \Delta A\ \Delta t\  \Delta \nu}\ \ \ ,
\label{e3.6}
\end{equation}
where the limit is taken for $\Delta A, \Delta t, \Delta \nu 
\rightarrow 0$. $F_\nu$ has the dimension 
[\,erg\,cm$^{-2}$\,s$^{-1}$\,Hz$^{-1}$\,] 
and can be derived from $I_\nu$ by summing over all projected 
contributions from the individual light rays in $\ve e_n'$ 
direction,
\begin{equation}
F_\nu \left(\ve r,\ve e_n ,t\right)=\oint I_\nu \left(
\ve r,\ve e_n',t\right) \ve e_n'\cdot \ve e_n\ \dtot\Omega'=
\int \limits_0^\pi \int \limits_0^{2 \pi} 
I_\nu\ {\underbrace {\cos\vartheta}_{\mu}} 
\ {\underbrace {\sin \vartheta\ \dtot\vartheta}_{-\dtot\mu}}\ \dtot\varphi\ \ \ , 
\label{e3.7} 
\end{equation} 
where $ \vartheta$ is measured from the $\ve e_n$
direction.  Here one takes into account that photons from $\ve e_n'$ 
see only a projected area $\Delta A  \ve e_n'\cdot \ve e_n$.
In most cases there is no $\varphi$ dependence, and 
\begin{equation}
F_\nu \left(\ve r,\ve e_n ,t\right) = 2 \pi \int \limits_{-1}^{+1}
I_{\nu}\ \mu\ \dtot\mu\ \ \ . 
\label{e3.8}
\end{equation}
Note that $F_\nu$ and $I_\nu$ are scalar quantities that depend on the
normal vector $\ve e_n$ of the
considered directed unit area. 
In an isotropic radiation field one has $F_\nu = 0$.

\subsection{Radiation Field, Level Populations, LTE and NLTE}
Consider the interior of a cavity which has been 
submerged for a long time in a gas of constant temperature $T$. 
All time-dependent processes have quieted down, and a final 
state has been reached, called {\it thermal equilibrium} (TE). 
It is described by four important relations: 
the {\it Planck function}, the {\it Boltzmann distribution}, the 
{\it Saha equation}, and the {\it Maxwell velocity distribution}.

\sss{a. Planck function}    
In TE the intensity is given by the {\it Planck function}:
\begin{equation}
 I _\nu =B_\nu \equiv {2h\nu^3 \over c^2}\  {1\over
{e^{h \nu/k T}-1}}\ \ \ ,  
\label{e3.9}
\end{equation}
which upon frequency integration gives
\begin{equation}
I= \int I_\nu \ \dtot\nu = B \equiv \int B_\nu \ \dtot\nu={\sigma\over
\pi} T^4\ \ \ . 
\label{e3.10}
\end{equation}
$B$ is called the {\it integrated Planck function}. Here 
$h=6.626\times 10^{-27}$ erg\,s is the Planck constant and
$\sigma=5.671\times 10^{-5}$ erg\,cm$^{-2}$\,s$^{-1}$\,K$^{-4}$  
the Stefan-Boltzmann constant.

\begin{figure}[t]
\includegraphics[height=2.5cm]{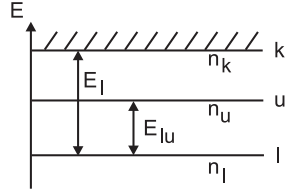}
\caption{Energy levels}
\label{huFig5}
\end{figure}

\sss{b. Boltzmann and Saha equations}
Consider the energy levels of the atoms or ions of a gas
(Fig.~\ref{huFig5}).  The number $n_l$ of atoms or ions per cm$^3$ in the
bound energy level $l$ is called the {\it population of level}
$l$.  $n_k$ is the population of the continuum, i.e., the number of
atoms or ions per cm$^3$ having their electron removed by ionization.
Bound levels in TE are described by the {\it Boltzmann distribution}:
\begin{equation}
{n_u\over{n_l}} ={g_u\over g_l} e^{-E_{lu}/k T}
={g_u\over g_l} e^{-h\nu_{lu}/k T}\ \ \ ,
\label{e3.11}
\end{equation}  
where $g_i$ are {\it statistical weights} (for hydrogen e.g.~$g_i=2i^2$) 
and $E_{l u}$ 
is the {\it energy difference} between the levels (for tabulated values 
of these quantities see Allen \citep{A73, C00}).  
Continuum levels in TE follow 
the {\it Saha equation}: 
\begin{equation}
 {n_k\ n_\mathrm{e} \over{n_l}}={2 u_k\over {g_l}}\left(
{2\pi m_\mathrm{e}kT\over{h^2 }}\right)^{3/2} e^{-E_l/k T}
\label{e3.12}
\end{equation}
where (see \citep{A73, C00}) $u_k$ is the {\it partition function}, 
$E_l$ the 
{\it ionization energy} from level $l$, $m_\mathrm{e} = 9.1094\times 10^{-28}$ g 
the mass of an electron, and $n_\mathrm{e}$ the number of electrons per cm$^3$.
Note that the number of particles per cm$^3$ is usually
called the number density of these particles.

\sss{c. Maxwell distribution}
In TE, atoms, ions, and electrons also obey the 
{\it Maxwell velocity distribution}:
\begin{equation}
{\dtot n\left(\vv\right)\over n}=4 \pi \vv^2 \left(m\over{2 \pi k T}
\right) ^{3/2} e^{-m\vv^2/{2kT}}\ \dtot \vv  
\label{e3.13}
\end{equation}  
Here $\dtot n(\vv)$ is the number of particles 
per cm$^3$ with mass $m$ and velocities in the interval $\vv$ to
$\vv+\dtot \vv$, while 
$n$ is total number of these particles per cm$^3$, irrespective of velocity.
\vskip2mm

\begin{table}[htb]
\begin{tabular}{llll}
\hline
  \tablehead{1}{l}{b}{TE}
 &\tablehead{1}{l}{b}{LTE}
 &\tablehead{1}{l}{b}{NLTE}
 &\tablehead{1}{l}{b}{Interplanetary}\\
\hline
$I_\nu =B_\nu $&$ I_\nu\neq B_\nu $&$ I_\nu\neq B_\nu  $&$ I_\nu\neq B_\nu $\\
Boltzmann   & Boltzmann       & $\neq$Boltzmann    & $\neq$Boltzmann\\
Saha        & Saha            & $\neq$Saha         & $\neq$Saha\\
Maxwell     & Maxwell         &  Maxwell           & $\neq$Maxwell\\
\hline
\end{tabular}
\caption{Various physical situations in stellar 
atmospheres defined by the successive break-down of relations valid in 
thermodynamic equilibrium.  Here the symbol $\neq$ means 
``not equal to'' or ``not valid''}
\label{huTab1}
\end{table}

\sss{d. Departures from thermodynamic equilibrium: LTE, NLTE } 
Deep in a star one has TE (Table 1).  Rising towards the surface, the
first concept that breaks down is the equality of the intensity and
the Planck function, while the Maxwell and Boltzmann distributions as
well as the Saha equation are still valid.  This is called {\it local
thermodynamic equilibrium} (LTE).  LTE holds roughly up to the
photosphere.  Rising further into the chromosphere and inner corona,
both the Boltzmann distribution and the Saha equation are no longer
valid.  This situation is called {\it Non-LTE} (NLTE). Here the
individual transition rates between the various energy levels must be
considered in detail.  In the low density parts of the corona the
temperatures of the different particle species become unequal, and
eventually in the interplanetary medium even the Maxwell distribution
breaks down.

\subsection{Absorption \& Emission Coefficients, Source Function}

\begin{figure}[h]
\includegraphics[height=3cm]{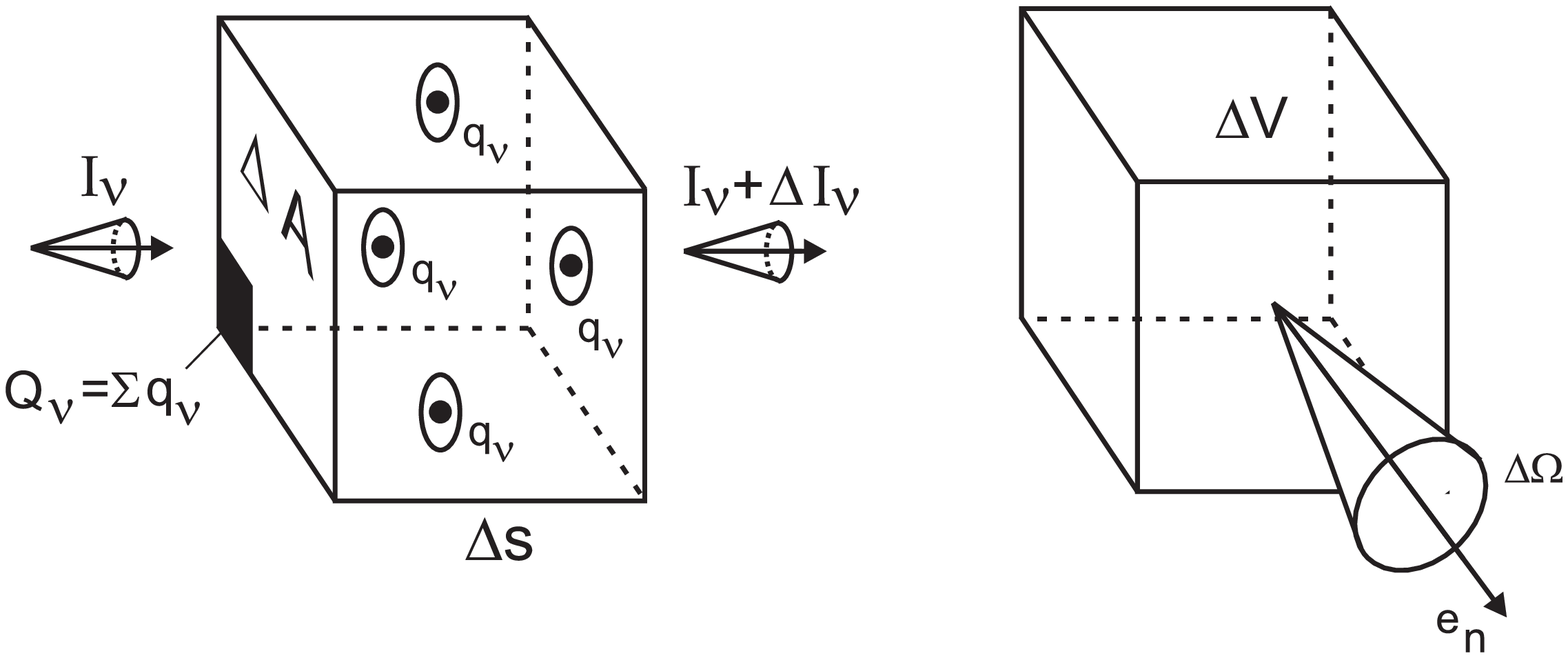}
\caption{Left: Absorption coefficient. Right: Emission coefficient}
\label{huFig6}
\end{figure}

\sss{a. Absorption coefficient}
Consider in Fig.~\ref{huFig6}, left, a ray of light with intensity
$I_\nu$, penetrating a box with surface area $\Delta A$ and
thickness $\Delta s$. The absorbing or scattering 
atoms in the box have a number density $n$ [\,cm$^{-3}$\,]
and cross section $q_\nu$ [\,cm$^2$\,].
The sum of all cross sections is $Q_\nu =\sum q_\nu$, 
thus:
\begin{equation}
{\Delta I_\nu \over{I_\nu}}= -{Q_\nu \over{\Delta A}}
=-{\sum q_\nu \over{\Delta A}}=-{q_\nu\ n\ \Delta A\ \Delta 
s\over \Delta A} = -\kappa_\nu\ \Delta s\ \ \ , 
\label{e3.14}
\end{equation} 
where
\begin{equation}
 \kappa_\nu \equiv q_\nu\ n\ \ \ 
\label{e3.15}
\end{equation}
is the {\it absorption coefficient} or {\it opacity} 
and has the 
dimension [\,cm$^{-1}$\,].  The function ${\kappa_\nu (T, p)}$ 
is available as a program package, e.g.\ in the program MULTI discussed below. 
It is often sufficient to consider only 
a gray (i.e., frequency averaged) Rosseland opacity: 
\begin{equation}
\overline {\kappa}= \left( \int \limits_0^\infty
{{1\over {\kappa_\nu}}{dB_\nu \over {\dtot T }} \dtot \nu} \over{
{\int \limits_0^\infty {dB_\nu \over{\dtot T}} \dtot \nu}} \right)^{-1}
\ \ \ . 
\label{e3.17}
\end{equation} 
At photospheric and low chromospheric temperatures the gray 
opacity $\overline \kappa$ can be approximated by the $H^-$
contribution:
\begin{equation}
 {\overline \kappa \over \rho}= 1.376\cdot 10^{-23}
p^{0.738} T ^5 \ \mathrm{cm}^2\,\mathrm{g}^{-1}\ \ \ . 
\label{e3.18}
\end{equation}

\sss{b. Emission coefficient and source function}
Let the volume $\Delta V$ (Fig.~\ref{huFig6}, right) emit photons of energy
$\Delta E$ in direction $\ve e_n$ into the solid
angle $\Delta \Omega$ per frequency interval
$\Delta \nu $ and per time interval $\Delta t$.  Then
\begin{equation}
\eta_\nu \left(\ve r,\ve e_n, t \right) \equiv
\lim {\Delta E \over \Delta V\ \Delta t\ \Delta\Omega\ \Delta \nu}
\label{e3.19}
\end{equation}
is the {\it emission coefficient} with the dimension 
[\,erg\,cm$^{-3}$\,s$^{-1}$\,sr$^{-1}$\,Hz$^{-1}$\,].  Here the limit is taken 
for $\Delta V$, $\Delta t$, $\Delta \Omega$, $\Delta \nu 
\rightarrow 0$. The {\it source function} is defined as
\begin{equation}
 S_\nu \left( \ve r,\ve e_n, t \right)\equiv 
{\eta _\nu \left(\ve r,\ve e_n,t \right) \over{\kappa_\nu
\left(\ve r,t \right)}}\ \ \ ,  
\label{e3.20}
\end{equation} 
which has the same dimension as the intensity.  In  TE 
the amount of energy 
absorbed, $\Delta E_A$, is exactly equal to the amount of
energy emitted, $\Delta E_E$
\[
\Delta E_A=\ B_\nu \ \kappa_\nu\ \Delta s\ \Delta A\ \Delta t\ 
\Delta \Omega\ \Delta \nu\ \ ;\ \ \  \Delta E_E=\ \eta_\nu\ 
\Delta A\ \Delta s\ \Delta
t \ \Delta \Omega\ \Delta \nu\ \ \ , 
\] 
which leads to {\it Kirchhoff's law}:
\begin{equation}
 S_\nu = B_\nu\ \ \ . 
\label{e3.21}
\end{equation} 
While in NLTE Kirchhoff's law no longer holds, it does 
hold in LTE because $S_\nu$, as 
we will see below, is essentially the population ratio, 
$n_u/n_l$, and from our definition of LTE, this ratio, 
same as in TE, obeys the Boltzmann distribution.

\subsection{Radiative Transfer}

\begin{figure}[h]
\includegraphics[height=4.5cm]{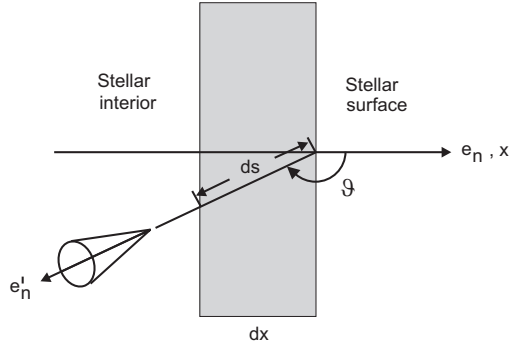}
\caption{Transfer equation}
\label{huFig7}
\end{figure}

\sss{a. Transfer equation}
Consider a gas layer in a stellar 
atmosphere (Fig.~\ref{huFig7}). Let $\ve e_n$ and the geometrical height $x$
point in the outward vertical direction.
Consider a light ray in an arbitrary 
direction $\ve e_n'$. The angle between $\ve e_n$
and $\ve e_n'$ is $\vartheta$. Let $s$ be the geometrical
distance along this light ray. From Eqs.~(\ref{e3.14}), (\ref{e3.19}),
and (\ref{e3.20}) we have
\begin{equation}
\dtot I_\nu =-I_\nu\ \kappa_\nu\ \dtot s + \eta_\nu\ \dtot s =
 -\kappa_\nu\ \left(I_\nu-S_\nu \right)\ \dtot s\ \ \ 
\label{e3.22}
\end{equation}
and
\begin{equation}
\dtot x=-\dtot s\ \cos \left( 180^\circ -\vartheta \right)=
\dtot s\ \cos \vartheta\ =  \mu\ \dtot s\ \ \ . 
\label{e3.23}
\end{equation}  

\par\noindent 
This gives the {\it radiative transfer equation}:
\begin{equation}
\mu\ {\dtot I_\nu \over \dtot x} =-\ \kappa_\nu\ \left(I_\nu- S_\nu 
\right)\ \ \ .  
\label{e3.24}
\end{equation}
We define the optical depth $\tau _\nu$:
\begin{equation}
 \tau_\nu \equiv - \int \limits_x^\infty \kappa_\nu\ \dtot x 
\ \ \ {\rm {\ or\ }}\ \ \  \dtot \tau _\nu=-\kappa_\nu\ \dtot x\ \ \ . 
\label{e3.25}
\end{equation} 
Using Eq.~(\ref{e3.25}), the transfer equation can be 
written in terms of the optical depth: 
\begin{equation}
 \mu\ {\dtot I_\nu \over{\dtot \tau_\nu}} = I_\nu-S_\nu\ \ \ . 
\label{e3.26}
\end{equation}

\sss{b. Formal solution of the transfer equation}
Multiplying Eq.~(\ref{e3.26}) with $e^{-\tau_\nu / \mu}$ one 
finds
\begin{equation}
 {\dtot \left( I_\nu e^{-\tau_\nu / \mu}\right) \over \dtot 
\tau_\nu /\mu} = -S_\nu e ^{-\tau_\nu / \mu}\ \ \ , 
\label{e3.27}
\end{equation}
which can be integrated between the limits $a$ and $b$:
\begin{equation}
 I_\nu e^{-\tau_\nu /\mu}{\Bigl\vert }_a^b\Bigr. 
=-\int \limits _a ^b S_\nu\ e^{-\tau_\nu ' / \mu}\ 
\dtot \tau_\nu '/\mu\ \ \ . 
\label{e3.28}
\end{equation} 
Boundary conditions:

\noindent i) outer boundary: No incoming radiation from outside 
the star 
\begin{equation}
I_\nu \left(0,\mu \right) = 0\ \ ,\ \ \ \ \mu <0 \ \ \ ,
\label{e3.29}
\end{equation}
ii) inner boundary: At the center of the star
the outgoing intensity is finite
\begin{equation}
 I_\nu \left(\tau _{\nu \infty},\mu \right)= {\rm {finite}}\ \ 
,
\ \ \ \ \mu \ge 0 \ \ \ .
\label{e3.30}
\end{equation} 
Consider the case $\mu <0$, take $b=\tau_\nu$ 
and $a=0$, and multiply Eq.~(\ref{e3.28}) with $e^{\tau_\nu / \mu}$:
\[
I_\nu \left(\tau_\nu,\mu \right)\ e ^{-\tau_\nu /\mu}\ e ^{\tau_\nu /\mu}
 -I_\nu \left(0,\mu \right)\ e^{\tau_\nu /\mu} =\hspace{7cm}
\]
\begin{equation}
\hspace{3cm} I_\nu \left(\tau_\nu,\mu \right) =-\int \limits _0^{\tau_\nu}
 S_\nu\ \left(\tau_\nu' \right)\ e ^{-\left( \tau_\nu '-\tau_\nu
\right)/\mu}\ \dtot\tau_\nu'/\mu \ \ \ , 
\label{e3.31}
\end{equation}
This is valid for {\it incoming radiation}, where 
$\mu <0$ and $\vartheta >\pi /2$.

Now consider the other case $\mu \geq 0$, take 
$b=\tau_{\nu \infty}$ very large
and $a= \tau_\nu$, and multiply Eq.~(\ref{e3.28}) with $-e^{\tau_\nu /\mu}$:
\[
 -I_\nu \left(\tau_{\nu \infty},\mu \right) 
\underbrace{e^{-\tau_{\nu \infty}/\mu}}_{=0}\ e^{\tau_\nu / \mu} +
I_\nu \left(\tau_\nu , \mu \right)\ e^{-\tau_\nu /\mu}\ 
e^{\tau_\nu /\mu} =\hspace{5cm} 
\]
\begin{equation}
\hspace{4cm} I_\nu \left(\tau_\nu, \mu \right) = \int \limits _{\tau_\nu}
^\infty\  S_\nu \left( \tau_\nu' \right)\ e^{-(\tau_\nu ' - \tau_\nu)/ 
\mu}\ \dtot\tau_\nu ' /\mu\ \ \ , 
\label{e3.32}
\end{equation} 
This is valid for {\it outgoing  radiation}, where 
$\mu \geq 0$ and $\vartheta < \pi /2$.

\subsection{Radiative Equilibrium, Eddington Approximation}

\sss{a. Radiative equilibrium}
Operate with $2\pi \int \limits _{-1} ^1 \dtot\mu $
on Eq.~(\ref{e3.24}) and use Eqs.~(\ref{e3.5}), (\ref{e3.8}):
\begin{equation}
{\dtot F_\nu \over \dtot x} = - 4\pi\ \kappa _\nu\ \left( J_\nu
-S_\nu \right)\ \ \ , 
\label{e3.37}
\end{equation} 
where  
\begin{equation}
 2 \pi\ \int \limits _{-1}^{+1}
S_\nu\ \dtot\mu = 2 \pi\ S_\nu \int \limits _{-1}
^{+1} \dtot\mu =4\pi\ S_\nu\ \ \ , 
\label{e3.38}
\end{equation}
if $S_\nu $ is assumed independent of $\mu $. 

If in a plane-parallel time-independent atmosphere the
energy transport is by radiation only, then $F= 
\int _0 ^\infty F _\nu\ \dtot\nu =$ const, i.e.\ there is a constant 
energy flux through all layers.  This condition is called
{\it radiative equilibrium}:
\begin{equation}
 {\dtot F\over \dtot x}= \int \limits _0 ^\infty {\dtot 
 F_\nu\over \dtot x} \dtot\nu = -4 \pi\ \int \limits _0 ^\infty \kappa_\nu
\ \left( J_\nu- S_\nu \right) \dtot \nu =0\ \ \ , 
\label{e3.39}
\end{equation} 
which for gray opacity $\kappa_\nu = 
\overline{\kappa}$ reduces to $J=S$ as $\overline{\kappa}$ can 
be taken out of the integration.
\vskip2mm

\begin{figure}[h]
\includegraphics[height=3cm]{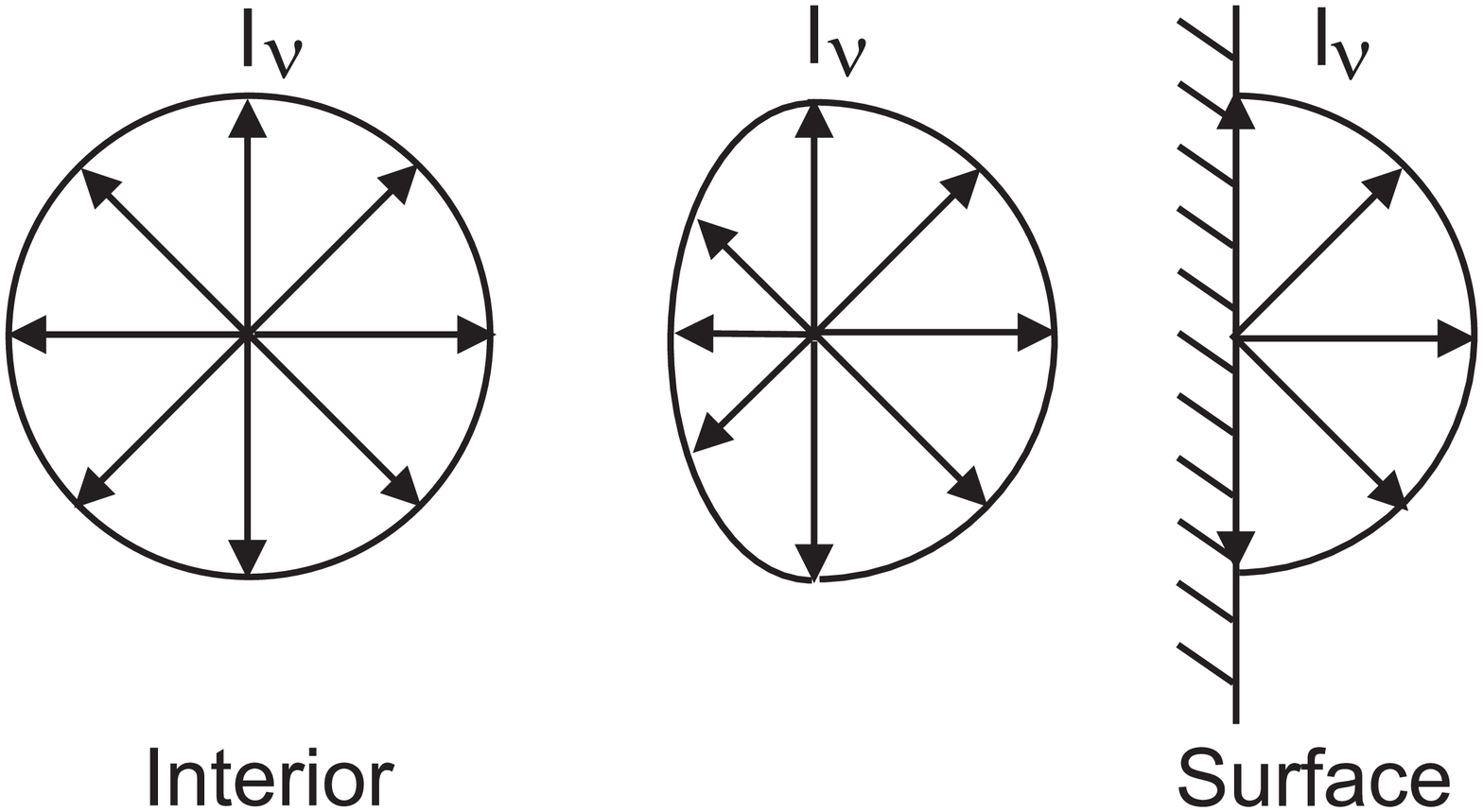}
\caption{Isotropy of the radiation field 
as function of depth in the star}
\label{huFig8}
\end{figure}

\sss{b. Eddington approximation}
Consider the behavior of the intensity
with depth (Fig.~\ref{huFig8}). 
If in deeper layers the intensity is approximately 
isotropic we get for the {\it Eddington K-integral}:
\begin{equation}
 K_\nu \equiv {1 \over 2} \int \limits _{-1}
^{+1} I_\nu \mu^2 \dtot \mu \simeq {1 \over 2} I_\nu 
\int \limits _{-1} ^{+1} \mu^2 \dtot \mu ={1\over 3} I_\nu\ \ \ . 
\label{e3.40}
\end{equation}
Similarly for the mean intensity at the same depth we obtain
\begin{equation}
 J_\nu ={1\over 2} \int \limits _{-1} ^{+1}
I_\nu \dtot \mu \simeq {1\over 2} I_\nu \int \limits
_{-1}^{+1} \dtot\mu =I_\nu\ \ \ . 
\label{e3.41}
\end{equation}
We thus find the {\it Eddington approximation}:
\begin{equation}
 K_\nu \simeq {1\over 3} J_\nu\ \ \ ,  
\label{e3.42}
\end{equation}
which gives relatively good results even in situations where
$I_\nu$ is fairly anisotropic.

\subsection{Gray Radiative Equilibrium Atmosphere}
Assume LTE, the Eddington approximation, radiative equilibrium,
and gray opacity; and integrate all equations over 
frequency $\nu$.  Then
\begin{equation}
 S=B \ \ ,\ \ \ \ K= {1\over 3 }\ J \ \ ,\ \ \ \ J=S\ \ \ , 
\label{e3.43}
\end{equation} 
from which we find with Eq.~(\ref{e3.10}):
\begin{equation}
 J = S = B = {\sigma \over \pi} T^4\ \ \ . 
\label{e3.44}
\end{equation}
Operate with ${1\over 2} \int \limits _{-1}
^{+1}\ \mu\ \dtot\mu$  on the transfer eq.~(\ref{e3.26}) and use Eqs.~(\ref{e3.8}), (\ref{e3.42}):
\begin{equation}
 {\dtot K\over \dtot \overline{\tau}} ={1 \over 3} {\dtot J\over{\dtot \overline{\tau}}}
= \underbrace { {1\over 2} \int \limits _{-1}^{+1} 
\mu\ I\ \dtot\mu}_{= F /4 \pi} - \underbrace{{1\over 2} S \int
\limits _{-1}^{+1} \mu\ \dtot\mu}_{=0} ={F\over 4 \pi}= 
{\rm{const.}} 
\label{e3.45}
\end{equation}
Multiply with 3 and integrate 
over $\overline \tau$:
\begin{equation}
 J= {\sigma \over \pi}\ T^4 = {3 F \over 4 \pi}  
\overline{\tau} + J(0)\ \ \ . 
\label{e3.46}
\end{equation}
At the surface the ingoing intensity is zero, so we need to integrate
only over $\mu > 0$ in Eqs.~(\ref{e3.5}) and (\ref{e3.8}):
\begin{equation}
 J \left( 0\right) = {1\over 2} \int \limits _0^1 B \left(0\right)
 \dtot \mu = {B \left(0 \right) \over 2}\ \ \ , \ \ \ 
 F \left( 0 \right) = 2 \pi \int \limits _0 ^1 B \left(0 \right)
 \mu \dtot \mu = \pi B \left( 0 \right)\ \ \ , 
\label{e3.48}
\end{equation}
thus
\begin{equation}
J \left(0 \right) = {F \left( 0\right) \over {2 \pi}}\ \ \ . 
\label{e3.49}
\end{equation} 
Let us define: 
\begin{equation}
 F= F\left( 0 \right) \equiv \sigma \ T_\mathrm{eff}^4\ \ \ ,
\label{e3.50}
\end{equation}
where $T_\mathrm{eff}$ is the {\it effective temperature}. 
Then we find from Eqs.~(\ref{e3.46}), (\ref{e3.49}), and (\ref{e3.50}) the so 
called {\it gray radiative equilibrium $T(\tau)$ relation}:
\begin{equation}
 T^4 = {3 \over 4}\ T_\mathrm{eff}^4\  \left( \overline \tau
+ {2\over 3} \right)\ \ \ . 
\label{e3.51}
\end{equation}
The relation between the optical and geometrical 
depths is obtained from Eq.~(\ref{e3.25}): 
\begin{equation}
\dtot \overline \tau =- \overline \kappa\  \dtot x\ \ \ . 
\label{e3.52}
\end{equation}
For a plane, static  (where 
the flow velocity $\ve v = 0$) atmosphere we find from Eq.~(\ref{e1.9})
the {\it equation of hydrostatic equilibrium}: 
\begin{equation}
\dtot p =-\rho\ g\ \dtot x\ \ \ . 
\label{e3.53}
\end{equation} 
Using for example the simple opacity law
(\ref{e3.18}), it is seen that Eqs.~(\ref{e3.51}), (\ref{e3.52}),
and (\ref{e3.53})
can be integrated as functions of $x$.  This allows to construct a
{\it static, plane, gray, radiative equilibrium atmosphere} 
if the two quantities $T_\mathrm{eff}$ and $g$ as well as suitable 
boundary conditions are given.  Together with Eq.~(\ref{e1.12}), $p=\rho
\Re T /\mu$, we have four equations for the four unknowns $T, 
p, \rho$, and $\overline \tau $.  Such radiative equilibrium atmospheres
are the starting atmosphere models for chromospheric wave calculations.

\subsection{Net Radiative Heating Rate, Radiative 
Heating Function}
The net radiative heating rate, $\Phi_\mathrm{R}$ [\,erg\,cm$^{-3}$\,s$^{-1}$\,],
(see Eq.~\ref{e1.11}) is the negative divergence
of the total radiative energy flux
$F= \int _0 ^\infty F_\nu\ \dtot\nu$,
thus in the 1D case we have 
\begin{equation}
 \Phi_\mathrm{R} = -{\dtot F \over \dtot x}\ \ \ .
\label{e3.54}
\end{equation}
Integrating Eq.~(\ref{e3.37}) over $\nu$ we find 
from  Eq.~(\ref{e3.54}) the radiative heating rate
\begin{equation}
 \Phi_\mathrm{R}=4\pi \int _0 ^ \infty \kappa_\nu \left(
J_\nu -S_\nu \right) \dtot \nu\ \ \ , 
\label{e3.56}
\end{equation}


\section{NLTE Thermodynamics and Radiation} 
As noted above (cf.\ Table 1), NLTE conditions prevail in the
chromosphere and corona, so that in these outer stellar regions the
individual transitions giving rise to lines and continua have to be
considered.  Chromospheres and coronae are regions where there is
mechanical heating and where departures from LTE are
important.  Starting with a slight departure in the upper photosphere,
NLTE becomes extreme in the high chromosphere and corona. It is
therefore necessary to review the transition rates and outline the
methods to treat the thermodynamics and radiation under NLTE
conditions.  General literature for this section includes \citep{M78,
R03, S02, A73, C00, SC85, C92, C95}.

\subsection{Transition Rates for Lines and Continua} 
Under NLTE conditions, there exists a well-defined kinetic temperature
$T$ that is determined by the {\it{Maxwell velocity distribution}},
but the {\it Boltzmann distribution} and the {\it Saha equation}
(Eqs.~\ref{e3.11}, \ref{e3.12}) are no longer valid.  Conservation of
particles requires that the population $n_m$ of level $m$ obeys
the {\it time-dependent statistical rate equation}
\begin{equation}
{\partial n _m\over \partial t} +\nabla\cdot n_m \ve v
=\sum_{j\ne m} n_j P_{jm} -  n_m \sum_{j\ne m} 
P_{mj}             
\ \ \ ,  
\label{e4.1}
\end{equation}
where $P_{ab}$ denotes the transition rates (= the number of transitions per sec) from level $a$ to level $b$.
In cases where populations adjust faster than the time scale over which
$n_m$ varies in hydrodynamic changes, the {\it statistical equilibrium equation} holds, 
\begin{equation}
 \sum_{j\ne m} n_j P_{jm} =\sum_{j\ne m} n_j (R_{jm} + C_{jm}) =  
n_m \sum_{j\ne m} P_{mj} =n_m \sum_{j\ne m} (R_{mj} + C_{mj})
\ \ \ ,
\label{e4.2}
\end{equation}  
where $R$ denotes radiative and $C$ collisional 
transition rates [\,cm$^{-3}$\,s$^{-1}$\,].  Here $R_\uparrow$ is the absorption rate,  
$R_\downarrow ^\mathrm{ind}$ the induced emission rate, $R_\downarrow ^\mathrm{sp}$ 
the spontaneous emission rate, $C_\uparrow$ the collisional excitation or ionization rate,
and $C_\downarrow$ the collisional deexcitation or recombination rate.

\begin{figure}[h] 
\includegraphics[height=4cm]{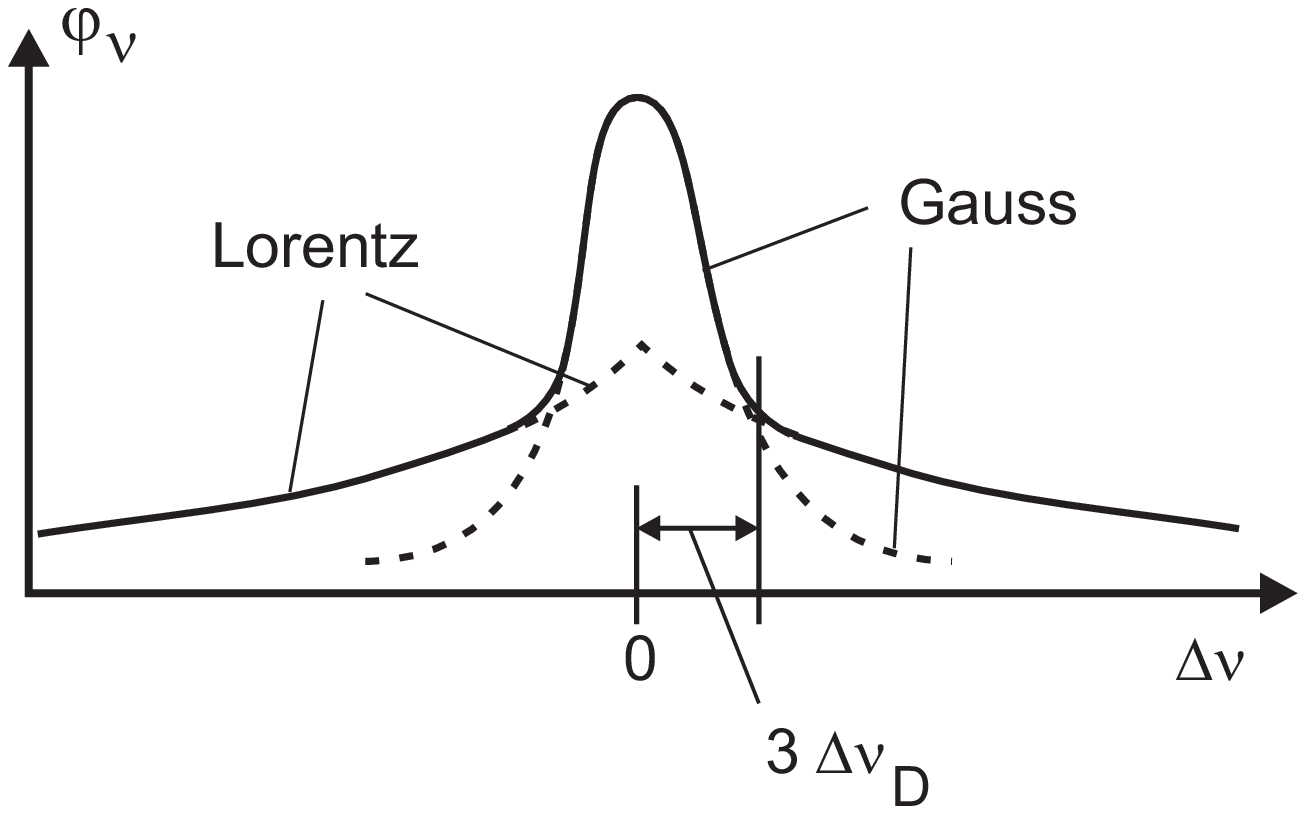}
\caption{The Voigt spectral line profile consists of folded Gauss 
and Lorentz profiles.} 
\label{huFig9}
\end{figure}

\sss{a. Lines} 
The \emph{radiative transition rates} 
between two bound energy levels, a lower level $l$ and 
an upper level $u$ (Fig.~\ref{huFig5}), are
\begin{equation}
R_\uparrow =n_l  R_{lu} =n_l B_{lu} \overline J_{lu}\ \ , \ \ 
R_\downarrow ^\mathrm{ind} =n_u B_{ul} \overline J_{lu} \ \ , \ \ 
R_\downarrow ^\mathrm{sp}=n_u A_{ul} \ \ ,
\label{e4.5}
\end{equation} 
where $A_{ul}$, $B_{lu}$, and $B_{ul}$ are the Einstein 
coefficients (tabulated e.g.\ by Allen \citep{A73, C00}).  $\overline J_{lu}$ is the mean 
intensity $J_\nu$, averaged over the line.  If $\ \varphi _\nu$ is the 
line profile, then 
\begin{equation}
\overline J_{lu} \equiv \int_{\Delta \nu}\varphi _\nu J_\nu \dtot \nu 
\ \ \rm{with}\ \  
\int_{\Delta \nu} \varphi_\nu \dtot \nu =1 
\ \ \ .
\label{e4.7}
\end{equation} 
Here the frequency integrals extend over the 
width ${\Delta \nu}$ of the line. In the above 
equations we have assumed that the emission and absorption 
profiles of the line are identical, that is, we assume {\it 
complete redistribution, CRD}. The line profile is 
usually given by the {\it Voigt profile} (see Fig.~\ref{huFig9}) 
\begin{equation}
 \varphi _\nu = {1\over\sqrt{\pi}\Delta\nu_\mathrm{D}}H(a,\vv)
\ \ \ ,
\label{e4.7a}
\end{equation}
where the {\it damping parameter} $a$ and the 
{\it normalized frequency separation} 
$\vv$ are given by
\begin{equation}
a = {\Gamma\over 4\pi\Delta\nu_\mathrm{D}}\ \ \ ,\ \ \
\vv = {\nu - \nu_0\over\Delta\nu_\mathrm{D}}
\ \ \ .
\label{e4.7b}
\end{equation}
Here $\nu_0$ is the line center frequency, $\Delta\nu_\mathrm{D}$ the 
{\it Doppler width} and $\Gamma$ the {\it damping constant}. 
In thermal 
equilibrium (TE), {\it{detailed balancing}} is valid, i.e.\ 
the radiation and collision processes individually balance 
each other,
\begin{equation}
R_\uparrow = R_\downarrow ^\mathrm{ind} + R_\downarrow ^\mathrm{sp} \ \ \ , \ \ \
 n_l B_{lu}\ B_\nu= n_u\ A_{ul}+n_u\ B_{ul}\ B_\nu 
\ \ \ . 
\label{e4.9} 
\end{equation} 
Dividing by $n_uB_{ul}$, solving 
for $B_\nu$, and 
using Eqs.~(\ref{e3.9}) and (\ref{e3.11}), one obtains
\begin{equation}
 B_\nu= {A_{ul} \over{B_{ul}}} {1 \over{{n_l \over{n_u}} 
{B_{lu} \over{B_{ul}}}-1}} = {A_{ul} \over 
{B_{ul}}} {1 \over{\left( {g_l\over {g_u}} e^ {{h \nu_{lu} 
\over{kT}}} \right) {B_{lu} \over{B_{ul}}}-1}} = 
{2h\nu^3 \over{c^2}}\ \ {1\over {e^{{h\nu \over{kT}}}-1}} 
\ \ \ ,
\label{e4.10}
\end{equation} 
and because $T$ is arbitrary 
\begin{equation}
 A_{ul}={2h\nu^3 \over{c^2}} B_{ul}  
 \ \ \ , 
\label{e4.11}
\end{equation} 
\begin{equation}
 g_l B_{lu} =g_u  B_{ul}
 \ \ \ . 
\label{e4.12}
\end{equation} 
From Eq.~(\ref{e4.5}) one sees that $A_{ul}$ has the dimension
[\,s$^{-1}$\,], while $B_{ul}$ and $B_{lu}$ have the dimension 
[\,cm$^2$\,sr\,Hz\,erg$^{-1}$\,].  Eqs.~(\ref{e4.11}) and 
(\ref{e4.12}) are relations between 
probabilities that involve only atomic parameters and do not 
depend on atomic level 
populations, they thus are {\it{also valid in NLTE}}.  Let us introduce an 
absorption cross section [\,cm$^2$\,] 
\begin{equation}
\alpha_{lu}(\nu) \equiv {\pi e^2\over m_\mathrm{e} c} f_l\varphi_\nu
\equiv B_{lu}{h\nu_{lu}\over 4 \pi}\varphi_\nu 
\ \ \ .
\label{e4.13}
\end{equation}
Here $e$ is the elementary charge, $m_\mathrm{e}$ the electron mass, and $f_l$ the 
oscillator strength.  Let us write the Boltzmann distribution 
(\ref{e3.11}) in the form
\begin{equation}
 {n_u^\star\over n_l^\star} = {g_u\over g_l} e^{-{h\nu_{lu} \over{k T}}}
\ \ \ ,
\label{e4.16}
\end{equation} 
where $g_i=2i^2$ and populations marked by a $\star$ 
indicate quantities in TE or LTE.  One finds
\begin{equation}
A_{ul} =
B_{lu}{2h\nu^3 \over{c^2}}  
{n_l^\star\over n_u^\star} e^{-{h\nu \over{k T}}}\ \ \ ,\ \ \ 
B_{ul} = 
B_{lu}{n_l^\star\over n_u^\star} e^{-{h\nu \over{k T}}}
\ \ \ .
\label{e4.16a}
\end{equation}
With this the radiative transition rates 
can be written
\begin{eqnarray}
R_\uparrow& = & 
n_l \int_{\Delta \nu} \alpha_{lu}(\nu) {4 \pi \over{h \nu}} 
J_\nu \dtot \nu  \equiv n_l R_{lu} \ ,
\label{e4.16b} \\
R^\mathrm{ind}_\downarrow& = &
n_u {n_l^\star \over n_u^\star}\int_{\Delta \nu}
\alpha_{lu} (\nu) 
{4 \pi \over{ h \nu}}J_\nu
e^{-{h\nu \over{k T}}} \dtot \nu \ ,
\label{e4.16c} \\
R^\mathrm{sp}_\downarrow& = &
n_u {n_l^\star \over n_u^\star}\int_{\Delta \nu}
\alpha_{lu} (\nu) 
{4 \pi \over{ h \nu}} {2h\nu^3 \over{c^2}} 
e^{-{h\nu \over{k T}}} \dtot \nu 
\ .
\label{e4.16d} 
\end{eqnarray}
Defining
\begin{equation}
G_{ul} \equiv {g_l\over g_u}
\ \ \ ,\ \ \ 
R_{ul}^\dagger \equiv \int_{\Delta \nu} \alpha_{lu} (\nu) 
{4 \pi \over{ h \nu}} \left({2h\nu^3 \over{c^2}} + J_\nu\right)
e^{-{h\nu \over{k T}}} \dtot \nu 
\ \ \ ,
\label{e4.16e} 
\end{equation}
the total radiative deexitation rate is given by
\begin{equation}
R_\downarrow  = R^\mathrm{sp}_\downarrow + R^\mathrm{ind}_\downarrow \equiv n_u R_{ul}
= n_u \int_{\Delta \nu} \alpha_{lu} (\nu) G_{ul}
{4 \pi \over{ h \nu}} \left({2h\nu^3 \over{c^2}} + J_\nu\right)
 \dtot \nu = n_u{n_l^\star \over n_u^\star} R_{ul}^\dagger
\ \ \ .
\label{e4.16g}
\end{equation} 
For the {\it collisional excitation and deexcitation rates} 
[\,cm$^{-3}$\,s$^{-1}$\,]
one has, respectively,
\begin{equation}
C_\uparrow =n_l\ C_{lu}=\ n_l\ n_\mathrm{e} \Omega _{lu} (T) 
\ \ \ , \ \ \ 
C_\downarrow = n_u\ C_{ul} \ = n_u\ n_\mathrm{e}  
\Omega_{ul} (T) 
\ \ \ . 
\label{e4.18}
\end{equation} 
The $\Omega$s are called collision cross sections. 
In TE, because of detailed balancing, one finds 
\begin{equation}
 C_{ul} = 
{n_l^\star \over{n_u}^\star} C_{lu} 
={n_l^\star \over{n_u}^\star} n_\mathrm{e} \Omega_{lu} (T)
={g_l \over{g_u}} e^{{h\nu_{ul} 
\over{kT}}} n_\mathrm{e} \Omega_{lu} (T) 
\ \ \ .
\label{e4.19}
\end{equation} 
This relation between the collision cross sections 
is {\it{also valid in NLTE}} because it involves only atomic parameters
and the Maxwell velocity distribution.

\sss{b. Continua}
The {\it radiative transition rates} [\,cm$^{-3}$\,s$^{-1}$\,] 
between a bound level $l$ and the continuum  
$k$ (Fig.~\ref{huFig5}) can be derived using similar arguments and
detailed balancing in each frequency interval:
\begin{equation}
R_\uparrow = n_l\  R_{lk}=n_l \int_{\nu_l } ^\infty 
\alpha_l \left(\nu \right) {4 \pi \over{h \nu}} 
J_\nu \dtot \nu  
\ \ \ , 
\label{e4.20}
\end{equation} 
\begin{equation}
 R_\downarrow ^\mathrm{ind}=n_k 
{n_l ^\star \over{n_k^\star}} \int_{\nu_l}^\infty \alpha_l 
\left(\nu \right) {4 \pi \over{h \nu}} J_\nu e^{- 
{h \nu \over{k T}}} \dtot \nu 
\ \ \ ,
\label{e4.26}
\end{equation} 
\begin{equation}
 R_\downarrow ^\mathrm{sp}=n_k{n_l^\star \over 
{n_k ^\star}} \int _{\nu_l} ^\infty \alpha_l \left(\nu 
\right) {4 \pi \over{ h \nu}} {2h\nu^3 \over{c^2}} 
e^{-{h\nu \over{k T}}} \dtot \nu 
\ \ \ .
\label{e4.27}
\end{equation} 
In TE and LTE the Saha equation (\ref{e3.12}) is written
\begin{equation}
{n_l^\star \over{n_k^\star}} =n_\mathrm{e} \left({h^2 \over{2 \pi m_\mathrm{e} kT}} 
\right)^{3/2} {g_l \over{2 u_k}} e^{{h \nu_l \over{k T}}} 
\ \ \ ,
\label{e4.28}
\end{equation} 
where $E_l=h\nu_l$ is the energy difference between level $l$ and the continuum.
$u_k$ in Eq. (\ref{e4.28}) is the partition function 
\begin{equation}
u_k = \sum_{i=1}^{i_\mathrm{max}}g_i e^{-{h(\nu_i-\nu_1) \over{k T}}}
\ \ \ ,
\label{e4.29}
\end{equation}
where the summation is carried out over the $i_\mathrm{max}$ bound levels of the next 
higher ionization stage.
It can be shown that Eqs.~(\ref{e4.26}) and (\ref{e4.27}) are also
valid in NLTE.
Defining
\begin{equation}
G_{kl} \equiv  {n_l^\star\over n_k^\star}e^{-{h\nu\over kT}}
\ \ \ ,\ \ \ 
R_{kl}^\dagger \equiv \int_{\nu_l}^\infty \alpha_{l} (\nu) 
{4 \pi \over{ h \nu}} \left({2h\nu^3 \over{c^2}} + J_\nu\right)
e^{-{h\nu \over{k T}}} \dtot \nu 
\ \ \ ,
\label{e4.29a} 
\end{equation}
the total radiative recombination rate is given by
\begin{equation}
R_\downarrow  = R_\downarrow^\mathrm{sp} + R_\downarrow^\mathrm{ind} 
\equiv n_k R_{kl}
=n_k \int_{\nu_l}^\infty\!
\alpha_l\left(\nu 
\right) G_{kl}{4 \pi \over{ h \nu}}\!\left({2h\nu^3 \over{c^2}} 
+J_\nu\right)\!  \dtot \nu 
= n_k{n_l^\star \over n_k ^\star} R_{kl}^\dagger\ \ \ .
\label{e4.30}
\end{equation} 
For the 
{\it collisional ionization and recombination rates} 
[\,cm$^{-3}$\,s$^{-1}$\,] one finds similarly 
\begin{equation}
C_\uparrow=n_l C_{lk}=n_l n_\mathrm{e} \Omega_{lk} \left(T \right) 
\ \ \ ,\ \ \ 
C_\downarrow = n_k C_{kl} =n_k n_\mathrm{e}^2 \overline\Omega_{kl} \left( 
T \right) \equiv n_k n_\mathrm{e}\Omega_{kl} \left( 
T, n_\mathrm{e} \right)                  
\ \ \ . 
\label{e4.32}
\end{equation} 
In TE, because of detailed balancing, one has 
\begin{equation}
C_\downarrow = n_k C_{kl}=n_k{n_l^\star \over{n_k^\star}} C_{lk} = 
n_k{n_l^ \star \over{n_k^ \star}}  n_\mathrm{e} \Omega _{lk} \left(T 
\right) 
\ \ \ ,
\label{e4.33}
\end{equation} 
with
${n_l^\star/{n_k ^\star}}$ given by Eq.~(\ref{e4.28}).  This relation is also valid in NLTE.
Partition functions, absorption and collision cross sections ($\alpha$'s and $\Omega$'s) can be found 
in Allen \citep{A73, C00} and references cited there.

\subsection{Line and Continuum Source Functions}
We now consider the transfer of radiation 
through a stellar gas.  After 
Eq.~(\ref{e3.20}) the source function is defined 
as the ratio of the emission and absorption coefficients.

\sss{a. Lines}
From Eqs.~(\ref{e3.19}) and (\ref{e4.5}), noting that the dimension 
of the transition rates is  
[\,cm$^{-3}$\,s$^{-1}$\,], one finds for the {\it line emission coefficient}
[\,erg\,cm$^{-3}$\,s$^{-1}$\,sr$^{-1}$\,Hz$^{-1}$\,]
\begin{equation}
\eta_\nu^\mathrm{line} = {\partial R_\downarrow^\mathrm{sp} \over{\partial \nu}} 
{h \nu \over{4 \pi}} =n_u A_{ul} 
{h \nu \over{4 \pi}} \varphi_\nu 
={2 h \nu^3\over c^2}\alpha_{lu}(\nu) n_u G_{ul} .
\label{e4.34}
\end{equation}
Similarly from 
Eqs.~(\ref{e3.14}) and (\ref{e4.5}), noting that
\begin{equation}
{\Delta I_\nu \over {\Delta s}} = -
I_\nu \kappa_\nu = - {\partial \left(R _\uparrow 
-R _\downarrow ^\mathrm{ind} \right) \over{ \partial \nu}} 
{h \nu \over{ 4 \pi}}
\ \ \ ,
\label{e4.35}
\end{equation} 
the {\it line opacity} [\,cm$^{-1}$\,] is given by
\begin{equation}
 \kappa_\nu^\mathrm{line}=\left(n_l B_{lu}-n_u B_{ul} \right) 
{h \nu \over{4 \pi}} \varphi _\nu= n_l B_{lu} 
\left(1-{n_u g_l \over{n_l g_u}} \right){h \nu \over 
{4 \pi}} \varphi _\nu
=\alpha_{lu}(\nu) (n_l - n_u G_{ul})  
\ .
\label{e4.36}
\end{equation} 
The absorption coefficient 
is defined for a total intensity change (which results from 
absorption minus induced emission).  The {\it line
source function} (see Eq.~(\ref{e3.20})) is
\begin{equation}
S_{lu}^\mathrm{line} \equiv {\eta _\nu ^\mathrm{line} \over 
{\kappa_\nu ^\mathrm{line}}} ={n_u A_{ul} \over {n_l B_{lu} 
-n_u B_{ul}}} 
={2h\nu^3 \over{c^2}}{1 \over 
{n_l g_u \over n_u g_l}-1} ={2h\nu^3 \over{c^2}}{1 \over
{b_l\over b_u}e^{{h \nu\over k T}}-1}  
\ \ \ ,
\label{e4.37}
\end{equation} 
where we have used the Boltzmann distribution.  
The {\it departure from LTE coefficient} $b_l$
is defined by
\begin{equation}
b_l \equiv {n_l\over n_l^\star} {n_k^\star \over n_k}
\ \ \ .
\label{e4.41}
\end{equation}
Note that always $b_k=1$.
It is seen that in LTE
\begin{equation}
S_{lu}^\mathrm{line} = B_{\nu}, \ \ \ {\text with}\ \ \ 
b_l = b_u = 1
\ \ \ .
\label{e4.37a}
\end{equation}
In deriving Eqs.~(\ref{e4.34}) and (\ref{e4.36}) we have assumed 
complete redistribution (CRD).

\sss{b. Continua} 
From 
Eqs.~(\ref{e3.19}) and  (\ref{e4.27}) we have similarly a 
{\it continuum emission coefficient} 
[\,erg\,cm$^{-3}$ s$^{-1}$\,sr$^{-1}$\,Hz$^{-1}$\,]
\begin{equation}
\eta_\nu^\mathrm{cont} ={\partial R _\downarrow ^\mathrm{sp} 
\over {\partial \nu}}{h \nu \over{4 \pi}} 
= n_k {n_l^\star \over{n_k^\star}} 
\alpha_l \left(\nu \right) {2 h \nu^3 \over{c^2}} 
e ^{-{h \nu \over{k T}}} 
={2 h \nu^3\over c^2}\alpha_{l}(\nu) n_k G_{kl} 
\ \ \ .
\label{e4.38}
\end{equation} 
Also, using Eqs.~(\ref{e4.20}), (\ref{e4.26}), and (\ref{e4.35}), we obtain the
{\it continuum opacity} [\,cm$^{-1}$\,]
\begin{equation}
\kappa_\nu ^\mathrm{cont} =\alpha_l \left( \nu \right) 
\left(n_l- n_k {n_l^\star \over{ n_k ^\star}} 
e^{-{h \nu \over{kT}}} \right)
=\alpha_{l}(\nu) (n_l - n_k G_{kl})  
\ \ \ . 
\label{e4.39}
\end{equation} 
The {\it continuum source function} is then given by 
\begin{equation}
 S_\nu ^\mathrm{cont}={\eta _\nu ^\mathrm{cont} \over \kappa_\nu ^\mathrm{cont}} 
={2h\nu^3 \over c^2} {1\over  {n_l\over 
n_l^\star} {n_k^\star \over n_k} e^{{h\nu \over kT}} 
-1}={2h\nu^3 \over{c^2}} {1\over  b_l
e^{{h\nu \over kT }}-1}
\ \ \  .
\label{e4.40}
\end{equation} 
Note that in LTE
\begin{equation}
S_\nu ^\mathrm{cont}=B_\nu,\ \ \ {\text with}\ \ \  b_l = 1
\ \ \ .
\label{e4.41a}
\end{equation}
Eqs.~(\ref{e4.37a}) and (\ref{e4.41a}) show the validity of 
Kirchhoff's law (\ref{e3.21}) in LTE, as stated already above.  
Taking 
the source function equal to the Planck function is used as primary definition
of LTE in some texts. For us here, the equality of the source and Planck functions (Eqs.~(\ref{e4.37a}) and
(\ref{e4.41a})) is the result of the way by which we defined LTE.

\subsection{Computation of the LTE and NLTE Level Populations}
Suppose the temperature $T$ and gas pressure $p$ are given along
with the chemical element abundance of the stellar gas. How can 
the number densities and level populations of the different
atoms and ions be computed?  Let
\begin{equation}
 n_{r,s,i}\ \ \ ,\ \ \ n_{s,i} = \sum_{r=1}^{r_\mathrm{max}}n_{r,s,i}
\ \ \ ,\ \ \ n_i = \sum_{s=1}^{s_\mathrm{max}}n_{s,i}
\ \ \ ,
\label{e4.55b}
\end{equation} 
be the number density
[\,cm$^{-3}$\,] of particles in energy level $r$ and ionization stage
$s$ of element $i$, the number density of particles of element $i$ 
in ionization stage $s$, and the total number density of particles 
of element $i$,
respectively. Here $r=1,\cdots, r_\mathrm{max}$, where 
$r_\mathrm{max}$ is the number 
of bound levels, and $s=1,\cdots, s_\mathrm{max}$, 
where $s_\mathrm{max}$ is the number of ionization stages.

\sss{a. LTE populations}
We first assume LTE.  In this case the Boltzmann distributions and
Saha equations are valid.  Summing Eq.~(\ref{e3.11}) over the bound 
levels $r$ we obtain 
the Boltzmann distributions with the partition functions
\begin{equation}
{n_{r,s,i}\over n_{s,i}} = {g_{r,s,i}\over u_{s,i}}\ 
e^{-{E_{r,s,i}\over k T}}
\ \ \ ,\ \ \ 
u_{s,i} = \sum_{r=1}^{r_\mathrm{max}} g_{r,s,i}
\ e^{-{(E_{r,s,i}-E_{1,s,i})\over k T}}
\ \ \ .
\label{e4.57}
\end{equation}
Similarly summing Eq.~(\ref{e3.12}) over all bound levels $l$ we find the Saha 
equations
\begin{equation}
{n_{s+1,i}\over n_{s,i}} = {1\over n_\mathrm{e}}{u_{s+1,i}\over u_{s,i}}\left({2 \pi m_\mathrm{e}k T\over 
h^2}\right)^{3/2}\ 
e^{-{E_{s-1,i}-E_{s,i}\over k T}}
\ \ \ .
\label{e4.58}
\end{equation}
One now proceeds as follows: 1. estimate $n_\mathrm{e}$. 
2. Compute the element densities
$n_i = \left({p/ k T} - n_\mathrm{e}\right){A_i/\sum_j A_j}$. 
3. Compute the $n_{s,i}$; for this we have $s_\mathrm{max}-1$ 
Eqs.~(\ref{e4.58}) and the third of the Eqs.~(\ref{e4.55b}).
Similarly compute $n_{r,s,i}$
using the $r_\mathrm{max}$ Eqs.~(\ref{e4.57}).
4. Evaluate the new electron density $n_\mathrm{e}$ using the 
equation of charge conservation
\begin{equation}
 n_\mathrm{e} =  \sum_i\sum_{s=2}^{s_\mathrm{max}} s\  n_{s,i}
 \ \ \ .
\label{e4.60}
\end{equation}
Going back to step 1, we could iteratively improve $n_\mathrm{e}$ 
until a converged solution is obtained.  One actually uses the much 
faster Newton-Raphson method by writing Eq.~(\ref{e4.60}) as 
$f(n_\mathrm{e})=0$.  If the true solution is written 
$n_\mathrm{e} = n_\mathrm{e}^0 +\delta n_\mathrm{e}$, 
where $n_\mathrm{e}^0$ is an estimate, we have $f(n_\mathrm{e}^0) + {\dtot 
f\over \dtot n_\mathrm{e}}\delta n_\mathrm{e} = 0$.  
The derivative can be evaluated
analytically.  Solving for $\delta n_\mathrm{e} = -f(n_\mathrm{e}^0)/{\dtot 
f\over \dtot n_\mathrm{e}}$ we get a new estimate $n_\mathrm{e}^0 +
\delta n_\mathrm{e}$, etc.
This Newton-Raphson iteration converges very fast if the 
initial estimate $n_\mathrm{e}^0$ is reasonable.

\sss{b. NLTE populations}
In NLTE the situation is different.  Here not only the Boltzmann
distributions and Saha equations are no longer valid, but also the
mean intensities $J_\nu$ are unknown.  Of the different methods to
solve the problem we discuss only the {\it complete linearization
method}. A computer code for this method, MULTI (\citet{SC85},
\citet{C86}, \citet{C92}) is available \citep{C95} and widely used
in the astrophysical community.

Here one also has a given
temperature $T$- and pressure $p$-distribution.  One starts by selecting an electron density $n_\mathrm{e}$-distribution 
and assuming level populations $n_i$.  Then the radiation fields
$J_\nu$ and the statistical rate equations are computed, which leads to
improved $n_i$.  In a Newton-Raphson scheme a system of equations, with a matrix ${\buildrel =\over W}$ operating on the 
vector of variations \{$\delta n_i$\} being equal to an error vector \{$E_i$\}, is inverted and
the new estimates $n_i+\delta n_i$ evaluated.  After convergence of this iteration the $n_\mathrm{e}$-distribution is
modified until the given $p$-distribution is obtained. Note that physical vectors are directed quantities in space,
mathematical vectors are simply arrays denoted by curly brackets \{ \}. 

Assume that for the $n^\mathrm{th}$ step of the iteration scheme we have the 
populations $n_i ^{\left(n\right)}$ and the transition processes 
$P_{ij} ^{\left(n \right)}$ and seek small corrections $\delta n_i$ and $\delta P_{ij}$.
From Eq.~(\ref{e4.2}),
written for level $i$, with the total number of levels $N_L=N+1$, we have
\begin{equation}
\left(n_i^{\left(n \right)} + \delta n_i ^{\left(n \right)}\right) \sum 
\limits _{j \ne i}^{N_L}\left( P_{ij} ^{\left(n \right)} + \delta P_{ij} 
^{\left( n\right)} \right) - \sum \limits _{j \ne i} ^{N_L} \left(n_j^{\left( n 
\right)} + \delta n _j ^{\left( n \right)} \right) \left( P_{ji} ^{ \left( n 
\right)} + \delta P_{ji} ^{\left( n \right) }\right) =0 
\ \ \ .
\label{e4.65}
\end{equation}
After expanding and neglecting second order terms we get
\begin{equation}
\delta n_i^{\left(n \right)} \sum \limits _{j \ne i} ^{N_L} P_{ij} ^{\left(n 
\right)} + n_i ^{\left(n \right)} \sum \limits _{j \ne i} ^{ N_L} \delta 
P_{ij} ^{\left(n \right)} - \sum \limits _{j \ne i} ^{N_L} \delta n_j 
^{\left(n \right)} P_{ji} ^{\left(n \right)} - \sum \limits _{j \ne i} ^{N_L} 
n_j^{\left(n \right)}\delta P_{ji}^{\left(n \right)} =E_i ^{\left(n \right)}
\ \ \ ,
\label{e4.66}
\end{equation} 
where $E_i ^{\left(n \right)}$ represents the zeroth order terms and
vanishes for the converged solution.
Our aim is to write $\delta P_{ij}, \delta P_{ji}$ in terms of $\delta n_i$,
and then to solve Eq.~(\ref{e4.66}) for $\delta n_i$.  With $\delta P_{ij}= \delta 
R_{ij} + \delta C_{ij}$, where $\delta C_{ij} =0$ and $\delta R_{ij} ^\mathrm{sp} =0$
because $n_\mathrm{e}$ and $T$ are given, we have 
\begin{eqnarray}
\delta P_{ij} = \delta R_{ij} & = &{1 \over 2} \int \limits _{-1} 
^{+1} \int 
\limits _{\Delta \nu} {4 \pi \over{h \nu}} \alpha _{ij} G_{ij} \delta I_{\nu 
\mu} \dtot \nu \dtot \mu\ \ ,\ \ i>j \nonumber\\
\ & = & {1 \over 2} \int \limits _ {-1} ^{+1} \int \limits 
_{\Delta \nu} {4 \pi \over{h \nu}} \alpha _{ij} \delta I _{\nu \mu} \dtot \nu \dtot \mu 
\ \ ,\ \ i<j  
\ \ \ .
\label{e4.67}
\end{eqnarray} 
$\alpha _{ij}$ is given by Eq.~(\ref{e4.13}) for lines, and $\alpha_{ik}= \alpha _i 
\left(\nu \right)$ from Eq.~(\ref{e4.20}) for continua.  The $G_{ij}$
are given by Eqs.~(\ref{e4.16e}) and
(\ref{e4.29a}), and the $\nu$-integration interval $\Delta \nu$ is 
either over the line width or from $\nu _i$ to infinity, 
depending on whether a line or a continuum transition is 
considered. 

\begin{figure}[t]
\includegraphics[height=5cm]{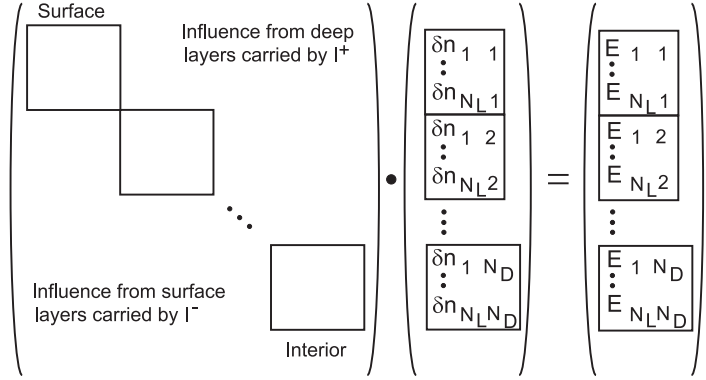}
\caption{Matrix equation in the complete linearization
  method as used by \citet{C86}}
\label{huFig10}
\end{figure}

\begin{figure}[h]
\includegraphics[height=5cm]{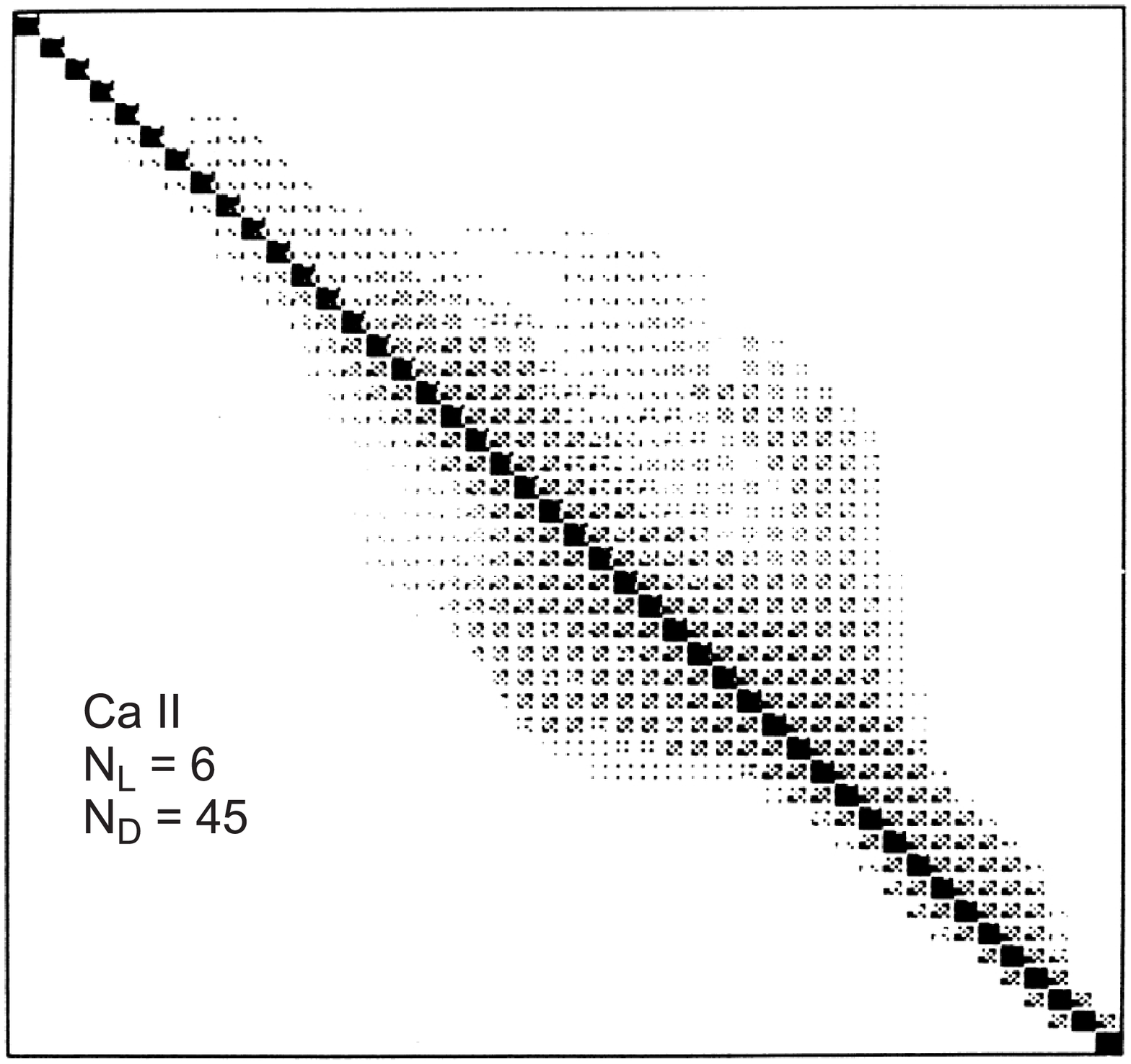}
\caption{Grand matrix ${\buildrel = \over W}$ for a 5 level + 
continuum Ca II calculation of the solar VAL81-C model \citep{VAL81}, 
after \citet{C86}.} 
\label{huFig11}
\end{figure}

The intensities $\delta I_{\nu \mu}$ are obtained from a linearization of the 
radiative transfer equations, where it is important that the in- and outgoing intensities 
at depth $d$ originate from other points of the atmosphere than depth $d$, depending on the
considered frequency.  This is done using a linear system in matrix form shown in Fig.~\ref{huFig10}.
With vectors $\delta \ve n=\{\delta n_{id}\}$, $\ve E= \{E_{id}\}$, 
where $i=1, \cdots , N_L$ is the energy level 
index and $d=1, \cdots , N_D$ the depth index, Eq.~(\ref{e4.66}) can be written
with a grand matrix  ${\buildrel = \over W}$, 
\begin{equation}
{\buildrel =\over W} \ \delta \ve n = \ve  E 
\ \ \ .
\label{e4.88}
\end{equation}%
As an example Fig.~\ref{huFig11} shows the non-vanishing elements of the grand
matrix ${\buildrel = \over W}$  for a 5 level + continuum Ca II
calculation.  The non-zero matrix elements are clustered along the main
diagonal, showing that the most important radiative interactions are of
intermediate range.  The off-diagonal matrix elements represent the
non-local contributions.  This band structure permits an efficient matrix
inversion. 

\subsection{Chromospheric Radiation Loss} 
From the procedure discussed above the net heating 
rates [\,erg\,cm$^{-3}$\,s$^{-1}$\,] for lines and continua
can be determined as
$\Phi_{lu},\ \Phi_{l} = h\nu \left(R_\uparrow - R_\downarrow\right)$:
\begin{equation}
\Phi_{lu} =  
n_l4 \pi\int_{\Delta\nu}\alpha_{lu}(\nu) J_\nu \dtot\nu\   
-n_u{n_l^\star\over n_u^\star}
4 \pi\int_{\Delta\nu}\alpha_{lu}(\nu)\left(
{2 h \nu^3\over c^2} + J_\nu\right)\ e^{-{h\nu\over k T}}\dtot\nu
\ \ \ ,
\label{e4.90}
\end{equation}
\begin{equation}
\Phi_{l} = 
n_l4 \pi\int_{\nu_l}^\infty\alpha_l(\nu) J_\nu \dtot\nu\   
-n_k{n_l^\star\over n_k^\star}
4 \pi\int_{\nu_l}^\infty\alpha_l(\nu)\left(
{2 h \nu^3\over c^2} + J_\nu\right)\ e^{-{h\nu\over k T}}\dtot\nu
\ \ \ .
\label{e4.91}
\end{equation}

\subsection{Coronal Radiation Loss}
In the tenuous coronal layers the temperature 
is high and the density is very low. 
Consider the energy levels of a typical multiply 
ionized coronal 
ion  with its large ionization $E_l$ and 
excitation $E_{lu}$ energies (Fig.~\ref{huFig5}). 
Because the photospheric radiation field $J_\nu$ 
does not provide photons with enough energy to excite a coronal 
ion,  
$R_\uparrow \approx 0$ and $R_\downarrow ^\mathrm{ind} \approx 0$.  In 
addition, as $n_\mathrm{e}$ is very small in the corona,
$C_\downarrow \approx 0$.  What remains from 
Eq.~(\ref{e4.2}) is 
\begin{equation}
C_\uparrow = R_\downarrow ^\mathrm{sp} 
\ \ \ ,
\label{e4.92}
\end{equation} 
which is called the {\it thin plasma approximation}.  From 
Eqs.~(\ref{e4.27}) and (\ref{e4.32}):
\begin{equation}
n_1n_\mathrm{e} \Omega_1 \left(T \right)=
n_k{n_1^\star\over n_k^\star}
\int_{\nu_1}^\infty\alpha_1(\nu){4 \pi\over h\nu}{2 h \nu^3\over c^2}
\ e^{-{h\nu\over k T}}\dtot\nu
\equiv n_k n_\mathrm{e} f\left(T \right)
\ \ ,
\label{e4.92a}
\end{equation} 
where we used the Saha eq.~(\ref{e4.28}),
\begin{equation}
{n_k^\star \over n_1^\star} ={1\over n_\mathrm{e}}\left({2 \pi m_\mathrm{e} kT\over h^2 } 
\right)^{3/2} {2 u_k \over g_1} e^{-{h \nu_1 \over{k T}}} 
\ \ \ .
\label{e4.92b}
\end{equation} 
and collected the temperature-dependent parts in some function
$f(T)$. This gives
\begin{equation}
{n_k \over{n_1}} ={\Omega_1 \left(T \right) \over{ 
f \left(T \right) }} \equiv g \left(T \right) 
\ \ \ .
\label{e4.93}
\end{equation} 
The function $g(T)$ can be tabulated (e.g.~\citet{JN78}). 
\begin{figure}
\includegraphics[height=4.2cm]{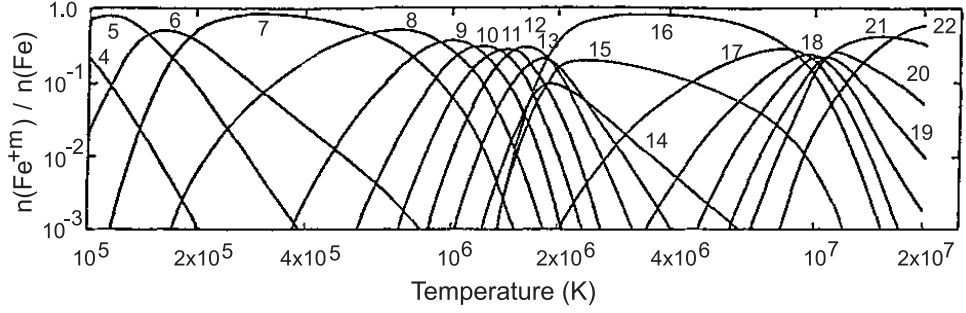}
\caption{Ionization ratios for Fe in the thin plasma 
approximation valid for the corona }
\label{huFig12} 
\end{figure}
Note that as lower bound level we took $l=1$ because only the ground 
level is significantly populated.  Eq.~(\ref{e4.93}) states that 
very differently from the Saha equation (\ref{e4.92b}) in LTE, which depends
also on $n_\mathrm{e}$, the coronal ionization ratio depends only on
$T$, due to the thin plasma approximation.  To illustrate this
dependence, Fig.~\ref{huFig12} shows the ionization ratios 
$n_{s,i}/\sum_s n_{s,i}$ for Fe in the coronal 
approximation.

\begin{figure}
\includegraphics[height=6cm]{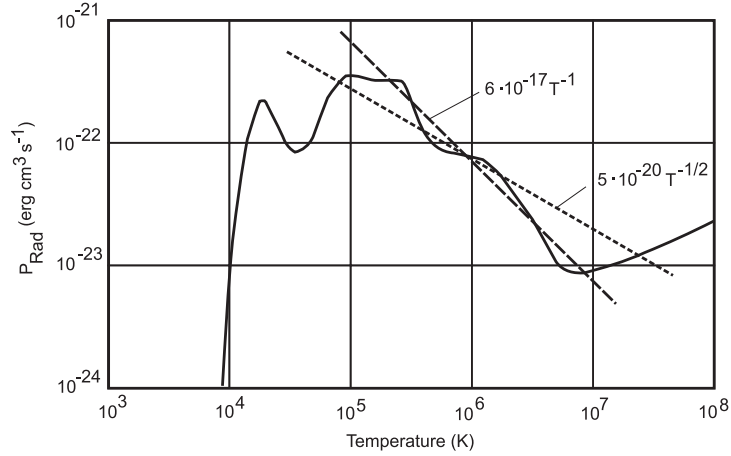}
\caption{Radiation function $P_\mathrm{Rad}$ as function of 
temperature and various numerical fits of this function, after 
\citet{MTW75}.}
\label{huFig13} 
\end{figure}

The coronal radiation loss due to lines is 
\begin{equation}
-\Phi_\mathrm{R}=4\pi \eta - \underbrace {4 \pi \kappa J}_{=0} 
=\sum h\nu \underbrace{n_u R_{u1}^\mathrm{sp}}_{n_1 C_{1u}} = 
\sum h \nu\ n_{Ion} 
n_\mathrm{e} \Omega \left(T \right)
\ \ \ ,   
\label{e4.93a}
\end{equation} 
where the sum is taken over all lines.  The total 
cooling rate includes also other processes and is given by
\begin{equation}
-\Phi_\mathrm{R} = n_\mathrm{H} n_\mathrm{e} P_\mathrm{Rad} \left(T \right)
 \ \ \ ,
\label{e4.94} 
\end{equation} 
where $P_\mathrm{Rad}(T)$ is a function shown in 
Fig.~\ref{huFig13}.  Similar functions were computed by numerous authors (e.g.,
\citep{CT69, MTW75, LL99}).
The descending part is often approximated by a simple power law.  
A dependence $\propto T^{-1/2}$, like
\begin{equation}
P_\mathrm{Rad} \approx 5\cdot 10^{-20}T^{-1/2}
\label{e4.95}
\end{equation}
(after \citep{MTW75}) is particularly useful as it allows to cast 
the coronal energy
balance in dimensionless form, thus making it possible to scale corona
models from one star to another (\citet{H84}).
Please note that Eq.~(\ref{e4.94}) defines only a radiative
{\it{cooling}} function, which will ultimately cool the gas down to  
temperatures $T\rightarrow 0$ in the absence of heating. 
In reality, if $T$ becomes small enough, 
{\it{radiative heating}} must also be considered, similarly as in 
$\Phi_\mathrm{R}=4 \pi \kappa \left(J-B \right)$.


\section{Simulations vs.\ Observations}

General references on the concepts discussed in this chapter include
\citep{C07, RU03, RFUM05, CS94}.

\subsection{Numerical Simulations}
In order to investigate the dynamics of the chromosphere
theoretically, it is necessary to solve the conservation equations for
mass (\ref{e1.25}), momentum (\ref{e1.27}), and energy (\ref{e1.39})
as functions of space and time for given initial and boundary values.
We also need the equation of state and other thermodynamic relations
necessary to close the system, Eqs.~(\ref{e1.12})\,--\,(\ref{e1.15}).
Simultaneously we must calculate the ionization rates, the electron
density, and the level population densities and radiative energy
source and sink terms, Eqs.~(\ref{e4.90}) and (\ref{e4.91}), for all
important spectral lines and continua.  As initial atmosphere one can
e.g.\ use a gray radiative equilibrium atmosphere, as derived above.
And as boundary conditions one typically specifies the velocity at the
lower end of the computational region and allows for waves to leave
the upper end with as little reflection as possible.

Ideally one would like to solve this system of partial differential
equations in three dimensions (3D), in order to be able to handle the
chromospheric structure.  Such 3D simulations exist (e.g.,
\citet{WFSLH04}), in some cases even including the effects of magnetic
fields (\citet{SWSF06,SVKWSF07,HCG07}) - but unfortunately the
currently available computer power does not yet permit the 3D
treatment of the full problem with all the physics that is important
in the chromosphere.  Radiative transport must either be treated in
gray LTE, or if NLTE effects are considered, they must be highly
simplified; moreover the spatial resolution of fine structures such as
shock waves is poor.  Nevertheless, these 3D simulations provide
impressive results about the structuring and dynamics of the
chromosphere (see also Steiner, this volume).

Another approach restricts itself to one spatial dimension (1D), but
solves the full NLTE radiation-hydrodynamics problem and calculates
the variation of line profiles, which can then be compared with
observations.  The two most sophisticated numerical codes to deal with
this problem are the one developed by M.~Carlsson and R.~F.~Stein
(briefly described in \cite{CS94, CS95, CS97, CS02}) and the one
developed by P.~Ulmschneider and collaborators already since the late
1970s (e.g., \citep{UKNB77}).  We will summarize the basic working of
the latter code, the most recent version of which is explained in 
detail in \citet{RU03}.

In order to solve a system of partial differential equations, one
basically replaces differentials by finite differences that connect
quantities of the known solution at the previous time level with those
at the new time level, and then one solves for the latter in order to
advance the solution in time.  

A special feature of the Ulmschneider-Rammacher approach is that before
this is done the differential equations are first transformed into
their characteristic form; i.e.\ one takes explicitly into account
that matter and hydrodynamic signals travel with speeds $\vv$
and $\vv\pm c$, respectively, thus defining the so-called
\emph{characteristics} in space-time (cf.~\citep{LL59, UKNB77}).  The
method of characteristics has a number of advantages: It is
numerically efficient \citep{HU78}, and it makes it easy to recognize
the formation of shocks (namely, when two neighboring characteristics
of the same kind intersect) and to handle the jump conditions at the
shock exactly - i.e., to take care of the continuity of mass,
momentum, and energy fluxes across the shock, while other variables
are allowed to change discontinuously.  Most other methods cannot
treat these discontinuities and use artificial or numerical viscosity
to spread out shocks over a certain height range, and then represent
each shock by a number of narrowly spaced grid points.  For these
reasons, characteristics methods with detailed shock handling are very
fast.  They are, however, best suited for 1D calculations, since the
necessary bookkeeping of characteristics gets prohibitively
complicated in 3D.

In this particular code, the solution is advanced in time in an
iterative process.  Suppose that all variables are known at some time
level $t$.  At the first time step these are the initial conditions
to be specified.  The radiative heating rates are first assumed to
remain constant, and a first guess is used for the values of the
hydrodynamic variables at a later time $t+\Delta t$.  This allows to
calculate the characteristics and the change of the variables along
these characteristics, leading to improved values of the hydrodynamic
variables in the next iteration.  This hydrodynamic iteration
converges after a few steps.  After the hydrodynamic variables at the
new time level are known, the population levels and radiative
intensities can be calculated, providing the radiative heating rates.
These are used in the next hydrodynamic iteration to further improve
the hydrodynamic variables, which in turn lead to improved population
levels and radiative heating rates, and so on.  If the radiative terms
or any other quantities are found to change too rapidly or too slowly during this
process, the time step $\Delta t$ is decreased or increased accordingly.

For reasons of computing time efficiency, the radiative part is
usually simplified during the simulation, and full line profiles
for diagnostic purposes are calculated with codes like MULTI only at
specific time steps.

A movie clip from an example calculation (\citet{R06}) shows how a
spectrum of waves moves through the atmosphere, steepening into shocks
that continue to grow, whereby occasionally larger ones catch up and
merge with smaller ones.  The associated emission is rather complex:
Even though a major contribution comes usually from behind large
shocks, because of the long-range interaction of chromospheric radiation
other parts of the atmosphere can also make significant contributions,
depending on the changing availability of emitters/absorbers and
photons in various parts of the highly dynamic atmosphere.  In the
movie clip \citep{R06} this is illustrated for two wavelengths in the
blue wing of Ca\,{\sc ii}\,K.

\subsection{Comparison with Observations}
Such 1D simulations have been very successful
(\citet{RU92}, \citet{CS92,CS97}) in explaining the \emph{Ca H and K
Bright Grain} phenomenon, a characteristic variation of the line
profiles of Ca\,{\sc ii}\,H and K that arises when strong acoustic
shocks traverse the mid-chromosphere.

Simulations in which the waves were injected according to photospheric
velocity measurements led to the surprising result that the average
temperature beyond a height of 500\,km (the canonical location of the
temperature minimum) did not increase, but continued to drop outward
(\citet{CS94,CS95}).  The emission behind strong shocks was found to
be so large that no general temperature rise was needed to generate
the chromospheric emission in lines such as Ca\,{\sc ii} H and K.
There has been some debate if this low average temperature is real or
caused by the neglect of high frequency waves, which are on principle
not observable due to their short wavelengths (\citet{KUA99},
\citet{C07}).  The problem could also be related to the 3D character
of wave propagation in the real solar chromosphere, which according to
\citet{URMK05} reduces the generation of very strong shocks by the
merging of weaker ones, as commonly found in 1D models, but rather
leads to a more continuous heating by a larger number of weak shocks.
Moreover, the meaning of an ``average'' temperature becomes
questionable in the presence of large temperature fluctuations
(\citet{CS95,RC05}).

Acoustic shock waves have long \citep{B46} been thought to be the main
heating agent of outer stellar atmospheres, or at least of the
nonmagnetic parts of stellar chromospheres (for a review see e.g.\ 
\citet{UM03}).  This view has been challenged recently, when Fossum
and Carlsson \citep{FC05, FC05b, FC06} used 1D simulations to
interpret the fluctuations observed by the \emph{TRACE} satellite in
UV continua formed in the upper photosphere.  They concluded that the
small observed fluctuations permit only an acoustic energy flux at
least an order of magnitude too small to balance the chromospheric
energy losses.  \citet{WSBR07} demonstrated, however, that this is
mostly due to the fact that the spatial resolution of \emph{TRACE} is
insufficient to resolve the fine-scaled lateral structuring found in
3D simulations.  And \citet{CRM07} argued that such a low acoustic
energy flux would be inconsistent with theoretical calculations of
sound generation (\citet{MRSU94}) and with the excellent agreement
between simulations and the measured Ca emission of the most inactive
stars.  Therefore, it appears that the acoustic heating theory is still valid
and cannot be considered dead at this time \citep{Twain97}.

\begin{figure}
\includegraphics[height=150mm]{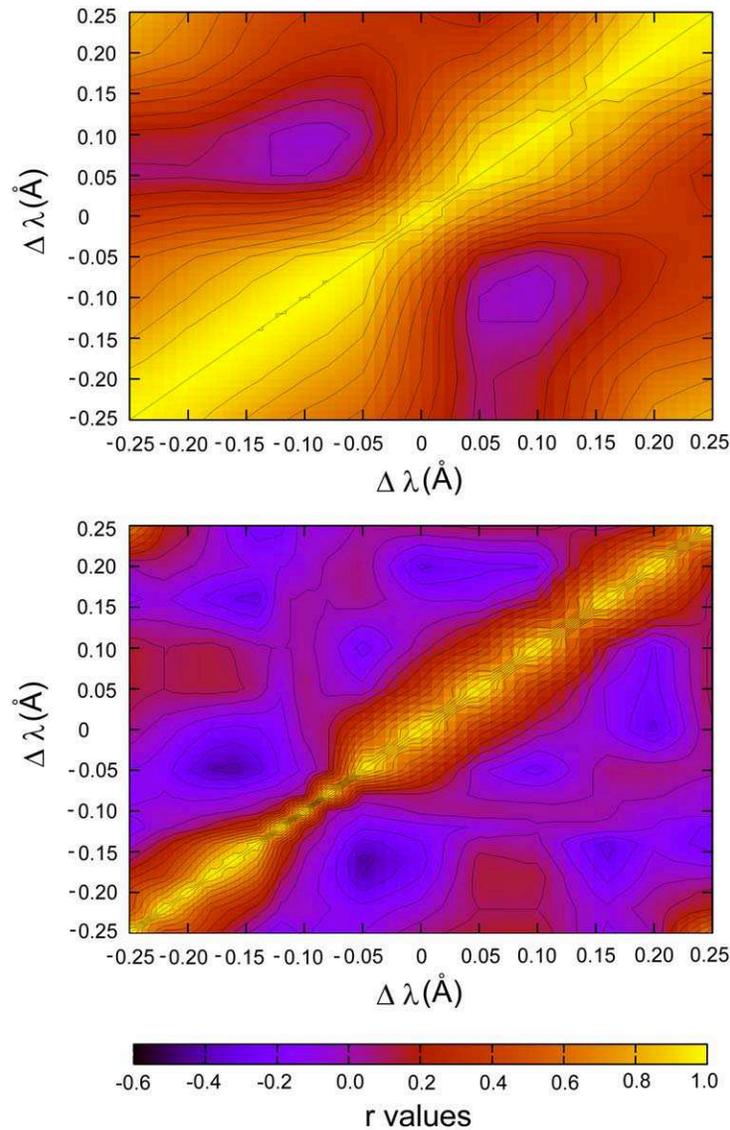}
\caption{Ca\,{\sc ii}\,H correlation matrices for VTT observations
(top panel) and for a 1D model simulation using a power spectrum with
total input energy flux of $5.0\times 10^7$ erg\,cm$^{-2}$\,s$^{-1}$
(bottom panel); from \citep{RSH07}.}
\label{huFig14} 
\end{figure}

Even though 1D simulations describe reasonably well the dynamics of
mostly vertically propagating strong shocks (Ca Bright Grains, as
discussed above), they fail to match the overall dynamics of the solar
chromosphere.  If the chromosphere were dominated by upward
propagating plane-parallel waves, the resulting perturbations in the
Ca\,{\sc ii} H and K lines would always start in the line wings (which
are formed in the photosphere) and then move inward towards the line
center (which is formed in the upper chromosphere).  Observations
\cite{RSHK07} show that this may happen indeed, however one often sees
large portions of the blue or red wing increase simultaneously.  To
quantify this behavior, W.~Rammacher \citep{RSH07} has introduced a
diagram that may serve as a ``fingerprint'' of the dynamics of the
chromosphere (Fig.~\ref{huFig14}).  It shows the correlation $r$ of the
intensity at any wavelength with the intensity at all other
wavelengths in the line profile.  If two parts of the atmosphere,
where two different wavelengths are formed, always vary
simultaneously, the correlation $r$ is 1.  Obviously this must
be true for the diagonal.  On the other hand, if the dynamics in the
two parts of the atmosphere are completely unrelated, the correlation
is $r=0$; whereas $r=-1$ means total anticorrelation, i.e., one part
of the atmosphere always brightens when the other darkens.
Observations (upper panel in Fig.\ \ref{huFig14}) show a much higher
degree of intensity correlation between different parts of the
atmosphere than 1D numerical simulations (lower panel).  The
difference to the observed fingerprint turned out to be large for all
numerical simulations, irrespective of the amount of wave energy and
the spectrum of wave frequencies used.  (According to \citep{R05} the
chromosphere depends less sensitively on the wave spectrum than on the
energy flux.)  Therefore, vertically propagating shock waves cannot
explain the observations.  \citet{RSH07} suggest that \emph{oblique}
shock fronts could explain the high observed correlations, because
they would lead to a simultaneous brightening of deeper and higher
parts of the atmosphere.  Such oblique shock fronts are commonly found
in 3D simulations (\citet{WFSLH04,WKBF05}); also \citet{URMK05} argued
theoretically that 3D propagation of shock waves must be important.

To summarize, the highly dynamic solar chromosphere (described in the
first chapter of this paper) calls for the simultaneous solution of
the radiation hydrodynamic equations under NLTE conditions (as
discussed in the next three chapters).  Modern computers allow their
full solution in 1D, while for 3D calculations, due to the present
lack of sufficient computational power, simplifications have to be
made.  Since 3D effects turned out to be important in the real solar
chromosphere (as discussed in this chapter), we will have to take
advantage of the best features of both types of calculations over the
next few years and combine them with the ever improving observational
capabilities in order to understand the dynamic solar chromosphere.


\begin{theacknowledgments}
We are grateful to Wolfgang Rammacher for comments on the manuscript
and for providing graphics materials.  RH thanks the Alexander von
Humboldt Foundation for travel support and the Indian Institute of
Astrophysics for the warm hospitality.
\end{theacknowledgments}

\newcommand\aj{Astron.~J.}
\newcommand\actaa{Acta Astron.}
\newcommand\araa{Ann.\ Rev.\ Astron.\ Astrophys.}
\newcommand\apj{Astrophys.~J.}
\newcommand\apjl{Astrophys.~J.}
\newcommand\apjs{Astrophys.~J.\ Suppl.}
\newcommand\ao{Appl.\ Opt.}
\newcommand\apss{Astrophys.\ \&Space Sci.}
\newcommand\aap{Astron.\ Astrophys.}
\newcommand\aapr{Astron.\ Astrophys.\ Rev.}
\newcommand\aaps{Astron.\ Astrophys.\ Suppl.}
\newcommand\azh{Astron.\ Zhurn.}
\newcommand\baas{Bull.\ Am.\ Astron.\ Soc.}
\newcommand\caa{Chinese Astron.\ Astrophys.}
\newcommand\cjaa{Chinese J.~Astron.\ Astrophys.}
\newcommand\icarus{Icarus}
\newcommand\jcap{J.~Cosmology Astropart.\ Phys.}
\newcommand\jrasc{JRASC}
\newcommand\memras{MmRAS}
\newcommand\mnras{Mon.\ Not.\ Roy.\ Astron.\ Soc.}
\newcommand\na{New Astron.}
\newcommand\nar{New Astron.~Rev.}
\newcommand\pra{Phys.\ Rev.~A}
\newcommand\prb{Phys.\ Rev.~B}
\newcommand\prc{Phys.\ Rev.~C}
\newcommand\prd{Phys.\ Rev.~D}
\newcommand\pre{Phys.\ Rev.~E}
\newcommand\prl{Phys.\ Rev.\ Lett.}
\newcommand\pasa{PASA}
\newcommand\pasp{PASP}
\newcommand\pasj{PASJ}
\newcommand\qjras{QJRAS}
\newcommand\rmxaa{Rev.\ Mexicana Astron.\ Astrofis.}
\newcommand\skytel{S\&T}
\newcommand\solphys{Solar Phys.}
\newcommand\sovast{Soviet Astron.}
\newcommand\ssr{Space Sci.\ Rev.}
\newcommand\zap{Zeitschr.\ Astrophys.}
\newcommand\nat{Nature}
\newcommand\iaucirc{IAU Circ.}
\newcommand\aplett{Astrophys.\ Lett.}
\newcommand\apspr{Astrophys.\ Space Phys.\ Res.}
\newcommand\bain{Bull.\ Astron.\ Inst.\ Netherlands}
\newcommand\fcp{Fund.\ Cosmic Phys.}
\newcommand\gca{Geochim.\ Cosmochim.\ Acta}
\newcommand\grl{Geophys.\ Res.\ Lett.}
\newcommand\jcp{J.~Chem.\ Phys.}
\newcommand\jgr{J.~Geophys.\ Res.}
\newcommand\jqsrt{J.~Quant.\ Spec.\ Radiat.\ Transf.}
\newcommand\memsai{Mem.\ Soc.\ Astron.\ Italiana}
\newcommand\nphysa{Nucl.\ Phys.~A}
\newcommand\physrep{Phys.\ Rep.}
\newcommand\physscr{Phys.\ Scr.}
\newcommand\planss{Planet.\ Space Sci.}
\newcommand\procspie{Proc.\ SPIE}
\let\astap=\aap
\let\apjlett=\apjl
\let\apjsupp=\apjs
\let\applopt=\ao


\bibliographystyle{aipproc}   


\bibliography{hu}

\hyphenation{Post-Script Sprin-ger}
\begin{thebibliography}{68}
\expandafter\ifx\csname natexlab\endcsname\relax\def\natexlab#1{#1}\fi
\providecommand{\enquote}[1]{``#1''}
\expandafter\ifx\csname url\endcsname\relax
  \def\url#1{\texttt{#1}}\fi
\expandafter\ifx\csname urlprefix\endcsname\relax\def\urlprefix{URL }\fi
\providecommand{\eprint}[2][]{\url{#2}}

\bibitem[{Vernazza} et~al.(1981)]{VAL81}
J.~E. {Vernazza}, E.~H. {Avrett}, and R.~{Loeser}, \emph{\apjs} \textbf{45},
  635--725 (1981).

\bibitem[{Fontenla} et~al.(1993)]{FAL93}
J.~M. {Fontenla}, E.~H. {Avrett}, and R.~{Loeser}, \emph{\apj} \textbf{406},
  319--345 (1993).

\bibitem[{Avrett}(2007)]{A07}
E.~H. {Avrett}, \enquote{{New Models of the Solar Chromosphere and Transition
  Region Determined from SUMER Observations},} in \emph{The Physics of
  Chromospheric Plasmas, ASP Conf.\ Ser.\ 368}, edited by P.~{Heinzl},
  I.~{Dorotovi\v c}, and R.~J. {Rutten}, San Francisco: ASP, 2007, pp. 81--91.

\bibitem[{Fontenla,} et~al.(2007)]{FBH07}
J.~M. {Fontenla,}, K.~S. {Balasubramaniam}, and J.~{Harder},
  \enquote{{Chromospheric Heating and Low-chromospheric Modeling},} in
  \emph{The Physics of Chromospheric Plasmas, ASP Conf.\ Ser.\ 368}, edited by
  P.~{Heinzl}, I.~{Dorotovi\v c}, and R.~J. {Rutten}, San Francisco: ASP, 2007,
  pp. 499--503.

\bibitem[{Solanki} and {Hammer}(2002)]{SH02}
S.~K. {Solanki}, and R.~{Hammer}, \enquote{{The Solar Atmosphere},} in
  \emph{The Century of Space Science}, edited by J.~A. {Bleeker}, J.~{Geiss},
  and M.~{Huber}, Berlin: Springer, 2002, pp. 1065--1088.

\bibitem[{Narain} and {Ulmschneider}(1990)]{NU90}
U.~{Narain}, and P.~{Ulmschneider}, \emph{Space Science Reviews} \textbf{54},
  377--445 (1990).

\bibitem[{Narain} and {Ulmschneider}(1996)]{NU96}
U.~{Narain}, and P.~{Ulmschneider}, \emph{Space Science Reviews} \textbf{75},
  453--509 (1996).

\bibitem[{Ulmschneider} and {Musielak}(2003)]{UM03}
P.~{Ulmschneider}, and Z.~{Musielak}, \enquote{{Mechanisms of Chromospheric and
  Coronal Heating},} in \emph{Current Theoretical Models and Future High
  Resolution Solar Observations: Preparing for ATST, ASP Conf.\ Ser.\ 286},
  edited by A.~A. {Pevtsov}, and H.~{Uitenbroek}, 2003, pp. 363--376.

\bibitem[{Kiepenheuer}(1953)]{K53}
K.~O. {Kiepenheuer}, \enquote{{Solar Activity},} in \emph{The Sun}, edited by
  G.~P. {Kuiper}, Chicago: Chicago University Press, 1953, pp. 322--465.

\bibitem[{Judge}(2006)]{J06}
P.~{Judge}, \enquote{{Observations of the Solar Chromosphere},} in \emph{Solar
  MHD: Theory and Observations, ASP Conf.\ Ser.\ 354}, edited by
  J.~{Leibacher}, R.~F. {Stein}, and H.~{Uitenbroek}, San Francisco: ASP, 2006,
  pp. 259--274.

\bibitem[{Hammer} and {Nesis}(2003)]{HN03}
R.~{Hammer}, and A.~{Nesis}, \enquote{{What Controls Spicule Velocities and
  Heights?},} in \emph{12th Cambridge Workshop on Cool Stars, Stellar Systems,
  and the Sun}, edited by A.~{Brown}, G.~M. {Harper}, and T.~R. {Ayres},
  \url{http://origins.colorado.edu/cs12/proceedings/poster/hammerxx.pdf}, 2003,
  pp. 613--618.

\bibitem[{Hammer} and {Nesis}(2005)]{HN05}
R.~{Hammer}, and A.~{Nesis}, \enquote{{A Metatheory about Spicules},} in
  \emph{13th Cambridge Workshop on Cool Stars, Stellar Systems and the Sun, ESA
  SP-560}, edited by F.~{Favata}, G.~{Hussain}, and B.~{Battrick}, 2005, pp.
  619--621.

\bibitem[{Rutten}(2007)]{R07}
R.~J. {Rutten}, \enquote{{Observing the Solar Chromosphere},} in \emph{The
  Physics of Chromospheric Plasmas, ASP Conf.\ Ser.\ 368}, edited by
  P.~{Heinzl}, I.~{Dorotovi\v c}, and R.~J. {Rutten}, San Francisco: ASP, 2007,
  pp. 27--48.

\bibitem[{De Pontieu} et~al.(2007)]{DePont07}
B.~{De Pontieu}, V.~H. {Hansteen}, L.~{Rouppe van der Voort}, M.~{van Noort},
  and M.~{Carlsson}, \enquote{{High Resolution Observations and Numerical
  Simulations of Chromospheric Fibrils and Mottles},} in \emph{{The Physics of
  Chromospheric Plasmas, ASP Conf.\ Ser.\ 368}}, edited by P.~{Heinzl},
  I.~{Dorotovi\v c}, and R.~J. {Rutten}, San Francisco: ASP, 2007, pp. 65--80.

\bibitem[{W\"oger}(2006{\natexlab{a}})]{W06}
F.~{W\"oger}, \emph{{High-resolution Observations of the Solar Photosphere and
  Chromosphere}}, {Ph.D.} dissertation, Univ. Freiburg (2006{\natexlab{a}}),
  \eprint{http://www.freidok.uni-freiburg.de/volltexte/2933/pdf/woeger_dissert%
ation.pdf}.

\bibitem[{W\"oger}(2006{\natexlab{b}})]{W06b}
F.~{W\"oger}, \url{http://www.kis.uni-freiburg.de/media/KS06/}
  (2006{\natexlab{b}}).

\bibitem[{Landau} and {Lifshitz}(1959)]{LL59}
L.~D. {Landau}, and E.~M. {Lifshitz}, \emph{{Fluid Mechanics}}, Oxford:
  Pergamon Press, 1959.

\bibitem[{Anderson} and {Athay}(1989)]{AA89}
L.~S. {Anderson}, and R.~G. {Athay}, \emph{\apj} \textbf{346}, 1010--1018
  (1989).

\bibitem[{Rammacher} et~al.(2007{\natexlab{a}})]{RSH07}
W.~{Rammacher}, W.~{Schmidt}, and R.~{Hammer}, \enquote{{Observations and
  Simulations of Solar Ca\,{\sc ii} H and Ca\,{\sc ii} 8662\,\AA\ Lines},} in
  \emph{The Physics of Chromospheric Plasmas, ASP Conf.\ Ser.\ 368}, edited by
  P.~{Heinzl}, I.~{Dorotovi\v c}, and R.~J. {Rutten}, San Francisco: ASP,
  2007{\natexlab{a}}, pp. 147--150.

\bibitem[{Rammacher} et~al.(2007{\natexlab{b}})]{RSHK07}
W.~{Rammacher}, W.~{Schmidt}, R.~{Hammer}, and W.~Kalkofen,
  \url{http://www.kis.uni-freiburg.de/media/KS06/} (2007{\natexlab{b}}).

\bibitem[{Hirschfelder} et~al.(1964)]{HCB64}
J.~O. {Hirschfelder}, C.~F. {Curtiss}, and R.~B. {Bird}, \emph{{Molecular
  Theory of Gases and Liquids}}, New York: Wiley, 1964.

\bibitem[{Spitzer}(1962)]{S62}
L.~{Spitzer}, \emph{{Physics of Fully Ionized Gases}}, New York: Interscience,
  2nd ed., 1962.

\bibitem[{Spiegel}(1969)]{S69}
M.~R. {Spiegel}, \emph{{Vector Analysis}}, New York: Schaum, 1969.

\bibitem[{Morse} and {Feshbach}(1953)]{MF53}
P.~{Morse}, and H.~{Feshbach}, \emph{{Methods of Theoretical Physics}}, New
  York: McGraw Hill, 1953.

\bibitem[{Mihalas}(1978)]{M78}
D.~{Mihalas}, \emph{{Stellar Atmospheres}}, San Francisco: W.~H.~Freeman and
  Co., 2nd ed., 1978.

\bibitem[{Rutten}(2003)]{R03}
R.~J. {Rutten}, \emph{{Radiative Transfer in Stellar Atmospheres}},
  \url{http://www.astro.uu.nl/~rutten/Astronomy_course.html}, 2003.

\bibitem[{Stix}(2002)]{S02}
M.~{Stix}, \emph{{The Sun: An Introduction}}, Berlin: Springer, 2nd ed., 2002.

\bibitem[{Seaquist}(2003)]{S03}
E.~{Seaquist}, \emph{{Radiation Processes}},
  \url{http://www.astro.utoronto.ca/~seaquist/radiation/}, 2003.

\bibitem[{Allen}(1973)]{A73}
C.~W. {Allen}, \emph{{Astrophysical Quantities}}, London: University of London,
  Athlone Press, 3rd ed., 1973.

\bibitem[{Cox}(2000)]{C00}
A.~N. {Cox}, \emph{{Allen's Astrophysical Quantities}}, New York: AIP Press;
  Springer, 4th ed., 2000.

\bibitem[{Scharmer} and {Carlsson}(1985)]{SC85}
G.~B. {Scharmer}, and M.~{Carlsson}, \emph{J.~Comp.\ Phys.} \textbf{59}, 56--80
  (1985).

\bibitem[{Carlsson}(1992)]{C92}
M.~{Carlsson}, \enquote{{The MULTI Non-LTE Program},} in \emph{Cool Stars,
  Stellar Systems, and the Sun, ASP Conf.\ Ser.\ 26}, edited by M.~S.
  {Giampapa}, and J.~A. {Bookbinder}, San Francisco: ASP, 1992, pp. 499--505.

\bibitem[{Carlsson}(1995)]{C95}
M.~{Carlsson}, \emph{{MULTI}}, \url{http://www.astro.uio.no/~matsc/mul22/},
  1995.

\bibitem[{Carlsson}(1986)]{C86}
M.~{Carlsson}, \emph{{A Computer Program for Solving Multi-level Non-LTE
  Radiative Transfer Problems in Moving or Static Atmospheres, Uppsala Astron.\
  Obs.\ Report No.\ 33}},
  \url{http://www.astro.uio.no/~matsc/mul22/report33.pdf}, 1986.

\bibitem[{Jain} and {Narain}(1978)]{JN78}
N.~K. {Jain}, and U.~{Narain}, \emph{\aaps} \textbf{31}, 1--9 (1978).

\bibitem[{Cox} and {Tucker}(1969)]{CT69}
D.~P. {Cox}, and W.~H. {Tucker}, \emph{\apj} \textbf{157}, 1157--1167 (1969).

\bibitem[{McWhirter} et~al.(1975)]{MTW75}
R.~W.~P. {McWhirter}, P.~C. {Thonemann}, and R.~{Wilson}, \emph{\aap}
  \textbf{40}, 63--73 (1975), {Erratum: {\it Astron.\ Astrophys.} {\bf 61}, 859
  (1977)}.

\bibitem[{Landi} and {Landini}(1999)]{LL99}
E.~{Landi}, and M.~{Landini}, \emph{\aap} \textbf{347}, 401--408 (1999).

\bibitem[{Hammer}(1984)]{H84}
R.~{Hammer}, \emph{\apj} \textbf{280}, 780--786 (1984).

\bibitem[{Carlsson}(2007)]{C07}
M.~{Carlsson}, \enquote{{Modeling the Solar Chromosphere},} in \emph{The
  Physics of Chromospheric Plasmas, ASP Conf.\ Ser.\ 368}, edited by
  P.~{Heinzl}, I.~{Dorotovi\v c}, and R.~J. {Rutten}, San Francisco: ASP, 2007,
  pp. 49--63.

\bibitem[{Rammacher} and {Ulmschneider}(2003)]{RU03}
W.~{Rammacher}, and P.~{Ulmschneider}, \emph{\apj} \textbf{589}, 988--1008
  (2003).

\bibitem[{Rammacher} et~al.(2005)]{RFUM05}
W.~{Rammacher}, D.~{Fawzy}, P.~{Ulmschneider}, and Z.~E. {Musielak},
  \emph{\apj} \textbf{631}, 1113--1119 (2005).

\bibitem[{Carlsson} and {Stein}(1994)]{CS94}
M.~{Carlsson}, and R.~F. {Stein}, \enquote{{Radiation Shock Dynamics in the
  Solar Chromosphere - Results of Numerical Simulations},} in
  \emph{Chromospheric Dynamics}, edited by M.~{Carlsson}, 1994, pp. 47--77.

\bibitem[{Wedemeyer} et~al.(2004)]{WFSLH04}
S.~{Wedemeyer}, B.~{Freytag}, M.~{Steffen}, H.-G. {Ludwig}, and H.~{Holweger},
  \emph{\aap} \textbf{414}, 1121--1137 (2004).

\bibitem[{Schaffenberger} et~al.(2006)]{SWSF06}
W.~{Schaffenberger}, S.~{Wedemeyer-B{\"o}hm}, O.~{Steiner}, and B.~{Freytag},
  \enquote{{Holistic MHD-Simulation from the Convection Zone to the
  Chromosphere},} in \emph{Solar MHD Theory and Observations: A High Spatial
  Resolution Perspective, ASP Conf.\ Ser.\ 354}, edited by J.~{Leibacher},
  R.~F. {Stein}, and H.~{Uitenbroek}, San Francisco: ASP, 2006, pp. 351--356.

\bibitem[{Steiner} et~al.(2007)]{SVKWSF07}
O.~{Steiner}, G.~{Vigeesh}, L.~{Krieger}, S.~{Wedemeyer-B{\"o}hm},
  W.~{Schaffenberger}, and B.~{Freytag}, \emph{Astronomische Nachrichten}
  \textbf{88}, 789--794 (2007), \eprint{astro-ph/0701029}.

\bibitem[{Hansteen} et~al.(2007)]{HCG07}
V.~H. {Hansteen}, M.~{Carlsson}, and B.~{Gudiksen}, \enquote{{3D Numerical
  Models of the Chromosphere, Transition Region, and Corona},} in \emph{The
  Physics of Chromospheric Plasmas, ASP Conf.\ Ser.\ 368}, edited by
  P.~{Heinzl}, I.~{Dorotovi\v c}, and R.~J. {Rutten}, San Francisco: ASP, 2007,
  pp. 107--114.

\bibitem[{Carlsson} and {Stein}(1995)]{CS95}
M.~{Carlsson}, and R.~F. {Stein}, \emph{\apjl} \textbf{440}, L29--L32 (1995).

\bibitem[{Carlsson} and {Stein}(1997)]{CS97}
M.~{Carlsson}, and R.~F. {Stein}, \emph{\apj} \textbf{481}, 500--514 (1997).

\bibitem[{Carlsson} and {Stein}(2002)]{CS02}
M.~{Carlsson}, and R.~F. {Stein}, \emph{\apj} \textbf{572}, 626--635 (2002).

\bibitem[{Ulmschneider} et~al.(1977)]{UKNB77}
P.~{Ulmschneider}, W.~{Kalkofen}, T.~{Nowak}, and U.~{Bohn}, \emph{\aap}
  \textbf{54}, 61--70 (1977).

\bibitem[{Hammer} and {Ulmschneider}(1978)]{HU78}
R.~{Hammer}, and P.~{Ulmschneider}, \emph{\aap} \textbf{65}, 273--277 (1978).

\bibitem[{Rammacher}(2007)]{R06}
W.~{Rammacher}, \url{http://www.kis.uni-freiburg.de/media/KS06/} (2007).

\bibitem[{Rammacher} and {Ulmschneider}(1992)]{RU92}
W.~{Rammacher}, and P.~{Ulmschneider}, \emph{\aap} \textbf{253}, 586--600
  (1992).

\bibitem[{Carlsson} and {Stein}(1992)]{CS92}
M.~{Carlsson}, and R.~F. {Stein}, \emph{\apjl} \textbf{397}, L59--L62 (1992).

\bibitem[{Kalkofen} et~al.(1999)]{KUA99}
W.~{Kalkofen}, P.~{Ulmschneider}, and E.~H. {Avrett}, \emph{\apjl}
  \textbf{521}, L141--L144 (1999).

\bibitem[{Ulmschneider} et~al.(2005)]{URMK05}
P.~{Ulmschneider}, W.~{Rammacher}, Z.~E. {Musielak}, and W.~{Kalkofen},
  \emph{\apjl} \textbf{631}, L155--L158 (2005).

\bibitem[{Rammacher} and {Cuntz}(2005)]{RC05}
W.~{Rammacher}, and M.~{Cuntz}, \emph{\aap} \textbf{438}, 721--726 (2005).

\bibitem[{Biermann}(1946)]{B46}
L.~{Biermann}, \emph{Naturwissenschaften} \textbf{33}, 118--119 (1946).

\bibitem[{Fossum} and {Carlsson}(2005{\natexlab{a}})]{FC05}
A.~{Fossum}, and M.~{Carlsson}, \emph{\apj} \textbf{625}, 556--562
  (2005{\natexlab{a}}).

\bibitem[{Fossum} and {Carlsson}(2005{\natexlab{b}})]{FC05b}
A.~{Fossum}, and M.~{Carlsson}, \emph{\nat} \textbf{435}, 919--921
  (2005{\natexlab{b}}).

\bibitem[{Fossum} and {Carlsson}(2006)]{FC06}
A.~{Fossum}, and M.~{Carlsson}, \emph{\apj} \textbf{646}, 579--592 (2006).

\bibitem[{Wedemeyer-B\"ohm} et~al.(2007)]{WSBR07}
S.~{Wedemeyer-B\"ohm}, O.~{Steiner}, J.~{Bruls}, and W.~{Rammacher},
  \enquote{{What is Heating the Quiet-Sun Chromosphere?},} in \emph{{The
  Physics of Chromospheric Plasmas, ASP Conf.\ Ser.\ 368}}, edited by
  P.~{Heinzl}, I.~{Dorotovi\v c}, and R.~J. {Rutten}, San Francisco: ASP, 2007,
  pp. 93--102.

\bibitem[{Cuntz} et~al.(2007)]{CRM07}
M.~{Cuntz}, W.~{Rammacher}, and Z.~E. {Musielak}, \emph{\apjl} \textbf{657},
  L57--L60 (2007).

\bibitem[{Musielak} et~al.(1994)]{MRSU94}
Z.~E. {Musielak}, R.~{Rosner}, R.~F. {Stein}, and P.~{Ulmschneider},
  \emph{\apj} \textbf{423}, 474--487 (1994).

\bibitem[{Mark Twain}(1897)]{Twain97}
{Mark Twain}, "{T}he report of my death was an exaggeration", New York Journal,
  June 2 (1897).

\bibitem[{Rammacher}(2005)]{R05}
W.~{Rammacher}, \enquote{{How Strong is the Dependence of the Solar
  Chromosphere upon the Convection Zone?},} in \emph{Chromospheric and Coronal
  Magnetic Fields, ESA SP-596}, edited by D.~E. {Innes}, A.~{Lagg}, and S.~A.
  {Solanki}, 2005.

\bibitem[{Wedemeyer-B{\"o}hm} et~al.(2005)]{WKBF05}
S.~{Wedemeyer-B{\"o}hm}, I.~{Kamp}, J.~{Bruls}, and B.~{Freytag}, \emph{\aap}
  \textbf{438}, 1043--1057 (2005).

\end{thebibliography}

\end{document}